\documentclass{elsarticle}

\usepackage{mathrsfs}
\usepackage{amssymb}
\usepackage{bbm}
\usepackage{amsfonts}
\usepackage{comment}
\usepackage[caption=false]{subfig}
\usepackage{graphicx}
\usepackage{psfrag}
\usepackage{amsmath}
\usepackage{bm}
\usepackage[colorlinks]{hyperref}
\usepackage[titletoc,title]{appendix}
\usepackage{algorithm}
\usepackage{algorithmic, algorithmic-fix}
\usepackage[margin=0.7in]{geometry}

\DeclareMathOperator*{\argmax}{arg\,max}

\journal{Journal of Computer Communications}

\begin{document}

\begin{frontmatter}

\title{Enhanced Energy-Efficient Downlink Resource Allocation in Green Non-Orthogonal Multiple Access Systems\tnoteref{mytitlenote}}
\tnotetext[mytitlenote]{This research was supported in part by the China NSFC Grant (61872248, U1736207), Guangdong NSF 2017A030312008, Shenzhen Science and Technology Foundation (No. JCYJ20170302140946299，JCYJ20170412110753954), Fok Ying-Tong Education Foundation for Young Teachers in the Higher Education Institutions of China(Grant No.161064)，GDUPS (2015) and GRCK2017082111300325. This work was also partially supported by the project "PCL Future Regional Network Facilities for Large-scale Experiments and Applications （PCL2018KP001)". Derrick Wing Kwan Ng was supported by the Australian Research Council's Discovery Early Career Researcher Award (DE170100137) and Discovery Project (DP190101363).}


\author[address1]{Rukhsana Ruby}
\ead{ruby@szu.edu.cn}
\author[address1]{Shuxin Zhong}
\author[address2]{Derrick Wing Kwan Ng}
\author[address1,address4]{Kaishun Wu\corref{mycorrespondingauthor}}
\cortext[mycorrespondingauthor]{Corresponding author}
\ead{wu@szu.edu.cn}
\author[address3]{Victor C.M. Leung}


\address[address1]{College of Computer Science and Software Engineering, Shenzhen University, Shenzhen, Guangdong, 518060 China}
\address[address2]{School of Electrical Engineering and Telecommunications, University of New South Wales, Australia}
\address[address3]{Department of Electrical and Computer Engineering, The University of British Columbia, Vancouver, Canada}
\address[address4]{PCL Research Center of Networks and Communications, Peng Cheng Laboratory, Shenzhen, China}

\begin{abstract}

Non-orthogonal multiple access (NOMA) technique has the ability to support multiple transmissions simultaneously using the same physical resource, the idea of which achieves much enhanced performance for a system compared to the conventional orthogonal multiple access (OMA) techniques. Despite numerous advantages, this phenomenon of NOMA technique can bring additional interference for the neighboring ultra-dense networks if the power consumption of the system is not properly optimized. While targeting on the green communication concept, in this paper, we propose an energy-efficient downlink resource allocation scheme for a NOMA-equipped cellular network. The objective of this work is to allocate subchannels and power of the base station among the users so that the overall energy efficiency is maximized. Since this problem is NP-hard, we attempt to find an elegant solution with reasonable complexity that provides good performance for some realistic applications. To this end, we decompose the problem into a subchannel allocation subproblem followed by a power loading subproblem that allocates power to each user's data stream on each of its allocated subchannels. We first employ a many-to-many matching model under the assumption of uniform power loading in order to obtain the solution of the first subproblem with reasonable performance. Once the the subchannel-user mapping information is known from the first solution, we propose a geometric programming (GP)-based power loading scheme upon approximating the energy efficiency of the system by a ratio of two posynomials. The techniques adopted for these subproblems better exploit the available multi-user diversity compared to the techniques used in an earlier work. Having observed the computational overhead of the GP-based power loading scheme, we also propose a suboptimal computationally-efficient algorithm for the power loading subproblem with a polynomial time complexity that provides reasonably good performance. Extensive simulation has been conducted to verify that our proposed solution schemes always outperform the existing work while consuming much less power at the base station.    
\end{abstract}


\end{frontmatter}

%\linenumbers

\section{Introduction}

% , which constrained the user allocation per subchannel to be at most two

% Furthermore, the proposed power loading scheme allocates power to each user's data stream on each of its allocated subchannels, rather than simply allocating power to each subchannel.

%  using the off-the-shelf computationally-intensive GP solvers

Next generation cellular networks \cite{QWu2017}\cite{MAgiwal2016} are envisioned to have some unique beneficial characteristics, among which ultra-dense deployment \cite{XGe2016} of network nodes is one of the most crucial paradigms. Ideally, ultra-dense deployment of nodes enhances the coverage and capacity of the system \cite{LZhou2014}. However, there are still some problems with this concept if the deployment and system parameters are not properly optimized. Since the number of transmitters in such networks is very large, this causes huge interference. Consequently, energy-efficient transmission is one of the important research endeavors in this arena. In order to obtain fruitful outcome in this context, designing an effective and efficient radio access technology \cite{TEdler2014} is one of the possible solutions. Non-orthogonal multiple access (NOMA) is considered as the most promising radio access technique for next generation wireless systems \cite{ZDing2014}\cite{KHIGUCHI2015}\cite{VWong2017}. Through experimentation and theoretical analysis \cite{Higuchi2013}\cite{YEndo2012}\cite{JUmehara2012}\cite{NOtao2012}, it is proved that the NOMA technique is able to provide enhanced performance in different sectors of wireless communication systems comparing with other orthogonal multiple access (OMA) techniques. Conceptually, power-domain NOMA \cite{SMRIslam2017} allows multiple users to occupy the same resource. This leads to additional interference for NOMA-equipped networks and their neighboring networks. Consequently, existing resource management techniques \cite{DFeng2013} in conventional networks, especially the energy-efficient ones, need to be revisited due to the incorporation of additional interference this new technology brings. 

Although the NOMA technique allows multiple users to be superimposed on the same frequency channel, due to the usage of practical, finite-length and possibly non-Gaussian codes, it is not an optimal design to assign large number users to the same channel. Consequently, dedicated spectrum of a system needs to be subdivided into multiple subchannels in order to support increased number of users~\cite{LDai2015}. At the same time, how to allocate these subchannels among the users in a multiplexed manner given the maximum allowable number of users that can utilize a subchannel simultaneously, is an important problem. While targeting on the maximization of the overall throughput, some research has been conducted on the downlink and uplink subchannel and power allocation in such NOMA systems. Based on some assumption of having constant power on the subchannels, existing works typically provide some heuristic solutions for the subchannel-user mapping problem. Once the subchannel-user mapping is done, in order to enhance the performance of the system further, different existing works have provided different schemes for the power allocation. For example, in~\cite{YSaito2013, ABenjebbour2013}, the authors use the fractional transmit power allocation technique among users and apply the equal power allocation technique across subchannels.~\cite{MRHojeij2015} uses the water filling-based approach for the power allocation, and in \cite{PParida2014}, the authors use difference of convex (DC) programming-based approach~\cite{NVucic2010} for the power allocation at both the user and subchannel levels. For uplink systems, there are some works as well, such as~\cite{KKumaran2003, MAlImari2015, MMollanoori2014}. In~\cite{MAlImari2015}, the authors used the iterative water filling idea~\cite{WeiYu2004} for both the subchannel-user mapping problem and their granular power allocation. On the other hand, in~\cite{MMollanoori2014}, the authors assumed that the resources are time slots instead of frequency channels. Although the work in~\cite{MAlImari2015} can ensure that users can be assigned to multiple subchannels, each user is assigned to at most one resource block via the scheme in~\cite{MMollanoori2014}. While the base station (BS) of all these works is assumed to be equipped with the half-duplex functionality, the authors in \cite{YSun2017} have proposed a joint subchannel and power allocation scheme for such a system based on the assumption that the BS is equipped with the full-duplex functionality to serve uplink and downlink users simultaneously.

With the increasing desire to have green communications in the recent years, reducing energy consumption has become the prime concern for researchers, and the fifth generation (5G) systems have also targeted energy efficiency as one of the major milestones to achieve~\cite{AAbrol2016}. Although the minimization of used power reported in \cite{LeiLei2016} could be one of the objectives of energy-efficient transmissions, the resultant solution of such a type of formulation is not spectrum-efficient. In order to achieve spectrum efficiency with the minimal power, the ideal objective of an energy-efficient transmission is to maximize the achievable bits of a channel under the unit Joule of power consumption~\cite{OJumira2012}. For a NOMA-equipped single cell, which has two users, an energy-efficient power allocation scheme is studied in~\cite{SHan2014}. In this work, the authors found the relationship between energy efficiency and spectrum efficiency under the total system power constraint. At the same time, energy efficiency is also studied for NOMA-equipped MIMO systems in~\cite{QSun2015}. The closest to ours and the most recent energy-efficient resource allocation scheme appeared in~\cite{FFang2016}. The system model that this paper considered is similar to ours, and our target is to enhance the performance of the proposed scheme in this paper further. This work has proposed an energy-efficient downlink subchannel and power allocation scheme for a cellular network under total power constraint at the BS with a number of drawbacks. For example, through their scheme, each user cannot be assigned to multiple subchannels, the concept of which fails to exploit the multi-user diversity of wireless systems~\cite{DavidTse2005}. Moreover, each subchannel can be assigned to at most two users, which also conflicts with the NOMA concept. Furthermore, they have used the DC programming-based approach in order to allocate power across all the subchannels and users of the system. Actually, through this power allocation, the entire power at the BS is used up, and hence the system fails to reach the optimal energy-efficient state. On the other hand, they also have utilized the DC programming-based approach in order to calculate the optimal instantaneous rate of a subchannel, which is an essential intermediate step of their subchannel-user mapping algorithm. However, the DC programming-based approach is appropriate only in this case because of provisioning at most two users to a subchannel. If the allowable number of users per subchannel is increased, the DC programming-based approach is not appropriate to solve this problem any more as the problem can no longer be expressed as a difference of convex functions under this realistic consideration.

%Typical definition of energy efficiency is achievable blocks of bits per unit Joule of power. Nonetheless, Shannon’s information capacity theorem illustrates that the two objectives of minimizing consumed energy and maximizing spectral efficiency are not achievable simultaneously. Note that under the consideration of fixed circuit power, there always exists an optimal point in the energy efficiency versus spectrum efficiency curve.

% Consequently, the problem is intractable as the authors in~\cite{FFang2016} have argued in their work. 

%  we have used many-to-many matching model~\cite{KHamidouche2014, ARoth1984} in order to overcome the drawbacks of the scheme in~\cite{FFang2016}. Through our proposed subchannel-user mapping algorithm based on many-to-many matching model, each user can have multiple subchannels and each subchannel can have more than two users as required. 

The contribution of this paper is the design of an energy-efficient downlink subchannel and power allocation scheme in a cellular network with enhanced performance compared to the work in~\cite{FFang2016}. Since this problem considers subchannels assignment which are associated with discrete variables in the formulated problem, the problem is NP-hard in general. Moreover, even if the subchannel assignment information is known, the power allocation problem is non-convex~\cite{DPBertsekas1999}. As a result, joint subchannel assignment and power allocation of this problem can be considered as an mixed integer non-linear programming (MINLP) problem. Considering the importance of the overall energy efficiency for such a system from the perspective of green communications, since the existing solution of~\cite{FFang2016} is approximate, the investigation of the better ones is required. Consequently, we find that decomposing the problem into a subchannel allocation problem followed by a power loading problem and then solving each problem individually while considering inter-dependency, is an elegant approach to solve this problem. In the first step, under the assumption that total power of the BS is subdivided equally among all the subchannels, we solve the subchannel-user mapping problem via a many-to-many matching model~\cite{KHamidouche2014, ARoth1984}. However, unlike the scheme in~\cite{FFang2016}, our proposed subchannel-user mapping scheme can utilize the multi-user diversity efficiently. On the other hand, one of the assumptions of the subchannel-users mapping algorithm in~\cite{FFang2016} is, at most $2$ users can be assigned to a subchannel. Because of this assumption, the power allocation problem of a subchannel is amenable to the DC programming-based approach. If a subchannel is assigned to more than $2$ users, the power allocation problem is no longer amenable to the DC programming-based approach. In contrast, our proposed scheme is general, and the number of users assigned to a subchannel is arbitrary. Our discovery to solve this problem via the geometric programming (GP)-based~\cite{Boyd2007, MChiang2005} approach is one of the specific contributions of this paper. Then, in the second step, once the subchannel-user mapping information are known, we adopt the GP-based approach in order to allocate power across all subchannel-user slots upon approximating the energy efficiency of system by a ratio of two posynomials. The DC programming-based approach in~\cite{FFang2016} allocates power across the subchannels while ignoring the detailed granularity of the user power level. Our proposed power loading scheme using the GP-based approach allocates power to each user's data stream on each of its allocated subchannels. Compared to the most relevant existing work~\cite{FFang2016}, the major contributions of the paper are summarized as follows. 

% solves this problem while taking detailed subchannel-user granularity into account

% Overall, while considering the overall energy efficiency, joint optimization of both subchannel assignment and power allocation is not tractable for this case. 

\begin{itemize}

\item For the subchannel-user mapping task, unlike the one in~\cite{FFang2016}, we have adopted a many-to-many matching model that exploits the multi-user diversity of wireless systems efficiently. Through this scheme, not only each subchannel can be used to serve multiple users, but also each user can utilize multiple subchannels. However, assigning more and more users to a subchannel cannot necessarily enhance the energy efficiency of the system further. Our proposed algorithm can smartly adapt user assignment to each subchannel given the information of some maximal allowable number of users to each subchannel.

\item While solving the subchannel-user mapping problem, it is required to determine the optimal instantaneous sum-rate of a subchannel in the intermediate decision making steps. When a subchannel can serve at most $2$ users, the approach in~\cite{FFang2016} is appropriate to determine the instantaneous sum-rate of a subchannel. However, in practice, having more than $2$ users on a subchannel can enhance the system performance. On the other hand, when a subchannel is assigned to more than $2$ users, the DC programming-based approach is no longer a suitable method to determine the optimal instantaneous sum-rate of that subchannel. This is because the objective function cannot be expressed as a difference of convex functions in this case. However, this problem is amenable to GP via the single condensation heuristic method after approximating the objective function of the problem by a ratio of two posynomials. Consequently, upon the approximation, we have adopted the single condensation method to determine the optimal instantaneous sum-rate of each subchannel while making the decision about each subchannel-user assignment.

\item Once the subchannel-user mapping information are known from the first step, the DC programming-based approach in~\cite{FFang2016} can allocate power optimally across the subchannels. However, this power allocation scheme ignores the detailed granularity of each subchannel-user slot, and the entire power of the BS is used up. While targeting on the optimality and considering the detailed granularity of each subchannel-user slot, we approximate the original problem via replacing the log function by the first term of its approximated series~\cite{RNave} to facilitate the problem solving via the GP-based approach. Upon approximating the objective function of the problem via a ratio of two posynomials, we adopt the GP-based single condensation heuristic to solve the approximated problem. The proposed scheme allocates power to each user's data stream on each of its allocated subchannels. As a result, the total required power, to reach the optimal overall energy-efficient state, is much less than the total power of the BS. From the perspective of green communications, this phenomenon of our power loading scheme is a big advantage over the DC programming-based approach. Although the proposed GP-based approach provides fine-grained power allocation across each subchannel and user with higher energy efficiency, this approach is computationally intensive with the growing number of users and subchannels due to the implementation limitation of the off-the-shelf GP solvers~\cite{cvx}. Therefore, based on the insights of the optimal solution, we also have proposed a computationally-efficient suboptimal solution for the fine-grained energy-efficient power loading problem, which has a polynomial time complexity.

\item Extensive simulation has been conducted in order to verify the effectiveness of our proposed resource allocation scheme. The results demonstrate that our scheme always outperforms the scheme in~\cite{FFang2016} under various realistic scenarios.

\end{itemize}   

% we adopt the GP-based approach to allocate power level across each subchannel-user slot. Since the problem is not directly amenable to GP, we apply some transformation on the original problem to make it amenable to GP. The log function of the original objective function is expanded through its approximated series. Then

The rest of the paper is organized as follows. Along with the background information and the description of the system, in Section~\ref{sec:sysmodel}, we formulate our energy-efficient resource allocation problem. The detailed solution approach is provided in Section~\ref{sec:sol}. Followed by the simulation methodology, we evaluate the performance of our proposed resource allocation schemes in Section~\ref{sec:eval}. Finally, Section~\ref{sec:concl} concludes the paper with some implication.

\section{System Model and Problem Formulation}
\label{sec:sysmodel}

In this paper, we consider a downlink scenario of a cellular network, which has one BS. Time is divided into frames, and the entire pre-assigned spectrum for the system is divided into $N$ subchannels in each frame. The resultant subchannels are the elements of a set, denoted by \textbf{N}. There are $M$ users in the system, and the corresponding set holding these users is denoted by \textbf{M}. Using the subchannels in set \textbf{N} as the transmission media, the BS transmits data to the users in set \textbf{M}. Both the BS and the users in the system are equipped with NOMA technologies. The BS transmits its data to a set of users using the superposition coding (SC) technique over a set of subchannels. Whereas, the receivers (i.e., the users) apply the SIC technique on each subchannel to decode the superimposed signals for extracting their own individual signal. However, before the downlink transmission operation, it is required to schedule subchannels and power across the users optimally so that the overall energy efficiency of the system is maximized. We assume that the scheduling scheme in the system is centralized, and the BS is appointed to conduct the entire scheduling operation. To develop this scheduling scheme, since the entire channel state information (CSI) of the system is required, the BS is aware of all these information. At the beginning of each time frame, all users send their CSI to the BS via some reliable control channels. The CSI of the subchannels in set \textbf{N} follows the block fading model~\cite{DavidTse2005}, where the CSI of each subchannel is constant for a time slot and varies in an i.i.d. manner from time slot to time slot. The BS has the maximal power constraint, denoted by $p^{\mbox{max}}$.

We assume that the BS assigns $M_n$ users to the $n$th subchannel, and the corresponding set holding these users is denoted by $\textbf{M}_n$. Given this, let denote the symbol transmitted by the BS on subchannel $n$ as $x_n$, and it is given by $\sum_{m \in \textbf{M}_n}\sqrt{p_m^n}{s_m}$. Here, $s_m$ is the modulated symbol of the $m$th user on subchannel $n$ and $p_m^n$ is the power level assigned to user $m$ on subchannel $n$. Consequently, the received signal of user $m$ on subchannel $n$ can be represented as

\[
y_m^n = \sqrt{p_m^n}h_m^ns_m + \displaystyle\sum_{i=1, i \neq m}\sqrt{p_i^n}h_m^ns_i + z_n,
\]

\noindent 
where $h_m^n$ is the channel gain of user $m$ on the $n$th subchannel. $z_n$ is the noise power over subchannel $n$, which follows Additive White Gaussian Noise (AWGN)~\cite{KMcClaning2000} distribution with mean zero and variance ${\sigma}_n^2$, i.e., $z_n \approx {\cal{CN}}(0, {\sigma}_n^2)$. The noise power of subchannel $n$ is statistically same for all users. In NOMA systems, each subchannel is shared by multiple users. Consequently, each user on subchannel $n$ receives its signal as well as the the interference signals from the other users that share the same subchannel. Let us denote the set holding the users that cause interference to user $m$ on subchannel $n$ as $\textbf{M}_n^m$. Given the information of set $\textbf{M}_n^m$, the received signal-to-interference-ratio (SINR) of the $m$th user on subchannel $n$ is given by

\begin{equation}
\mbox{SINR}_m^n = \frac{p_m^n|h_m^n|^2}{{\sigma}_n^2 + \displaystyle\sum_{i \in \textbf{M}_n^m}p_i^n|h_m^n|^2} = \frac{p_m^n{g_m^n}}{{\sigma}_n^2 + \displaystyle\sum_{i \in \textbf{M}_n^m}p_i^ng_m^n},
\end{equation}

\noindent
where ${\sigma}_n^2 = E[|z_n|^2]$ is the noise power on subchannel $n$, and $g_m^n = |h_m^n|^2$ represents the power gain of the $m$th user on subchannel $n$. 

%Because of equipping NOMA technologies, the objective of the SIC process at the receivers is to reduce the interference from the other users on the same subchannel. {\color{red}In downlink systems, since the same transmitter (i.e., the BS) transmits data to the users in set $\textbf{M}_n$ on subchannel $n$, the optimal decoding order of the users and their power assignment are coupled to each other, and the resultant outcome of this coupled problem is a crucial factor in extracting the signal of each individual user. Unlike uplink systems~\cite{MMollanoori2014}, if a larger power level is assigned to a user with larger gain and a lower power level is assigned to a user with lower gain, it may be possible that the users with lower gain cannot decode their signals due to the insufficient power level that they are assigned with. However, if these users cannot decode their signals, this is not a fair attitude from the overall performance point of view although the power allocation in the other way is throughput optimal. Therefore, in order to reduce the overall computational complexity, similar to~\cite{FFang2016}, the users with larger gain are assigned with lower power level compared to the users with lower gain. Following this decoding order, the set of users that cause interference to user $m$ on subchannel $n$ is given by $\textbf{M}_n^m = \{m'|_{g_{m'}^n > g_m^n}, m' \in \textbf{M}_n\}$. Consequently, the sum-rate of subchannel $n$, denoted by $R_n(\textbf{M}_n, \{p_m^n\})$, can be obtained by Shannon's capacity formula~\cite{HMichiel2001}.} 

Since the NOMA users are equipped with the SIC technique, the way a NOMA user retrieves information depends on the order of the allocated power level among all the NOMA users in the corresponding subchannel. In other words, a particular NOMA user can decode the information of other users that are assigned to larger power level compared to itself. On the other hand, this user can decode its own information considering the power level of other remaining users (that are assigned to lower power level compared to itself) as noise. Therefore, if the order of the power level assigned to the NOMA users are known, the decoding order is straightforward for a particular NOMA user and vice versa. It is clear that the optimal decoding order of the NOMA users and their power assignment are coupled to each other and is not a straightforward problem. However, in order to reduced the complexity of the proposed solution scheme in this paper, we intend to fix the order of the power level among the NOMA users while solving the entire energy efficiency problem. Ideally, in order to maximize the energy efficiency of the system, the optimal power allocation among the NOMA users of a subchannel should follow the water filling-based approach. On the other hand, when the fairness of the system is considered, the order of the ideal power allocation should be in-between the water filling-based approach and the opposite of the water filling one. If we would fix the order of the power level among the NOMA users according to the water filling norm, it is very likely that an ideal power loading scheme\footnote{The proposed power loading scheme in this paper is particularly optimal for the applications that work in the low SNR regime.} assigns very low power to the users with worse channel condition. In this case, those users (with worse channel condition) may not be able to decode their information or this is not a fair attitude to those users. Therefore, while considering the fairness of the system, we have adopted the opposite norm of the water filling technique so that the users with worse channel have fair provision. In other words, in order to reduce the overall computational complexity of the solution schemes and considering the fairness of the system, similar to~\cite{FFang2016}, the users with larger gain are assigned to lower power level compared to the users with lower gain. As a result, following the norm of the SIC technique, the set of users that cause interference to user $m$ on subchannel $n$ is given by $\textbf{M}_n^m = \{m'|_{g_{m'}^n > g_m^n}, m' \in \textbf{M}_n\}$. Consequently, the sum-rate of subchannel $n$, denoted by $R_n(\textbf{M}_n, \{p_m^n\})$, can be obtained by the Shannon's capacity formula~\cite{HMichiel2001}.

\[
R_n(\textbf{M}_n, \{p_m^n\}) = \displaystyle\sum_{m \in \textbf{M}_n}\mbox{log}_2\left(1 + \mbox{SINR}_m^n\right).
\]

If the circuit power consumption on subchannel $n$ is denoted by $p_c$, the energy efficiency of subchannel $n$ is given by

\[
E_n = \frac{R_n(\textbf{M}_n, \{p_m^n\})}{p_c + \sum_{m \in \textbf{M}_n}p_m^n}.
\]

%\noindent
%which is actually the transmitted data bits per unit of power level.  

In this work, given the power constraint of the BS, our objective is to allocate the subchannels in set \textbf{N} across all the users in set \textbf{M} so that the aggregate energy efficiency of all subchannels, $\sum_{n \in \textbf{N}}E_n$, is maximized. Clearly, this is an optimization problem. To formulate this problem, we define binary variables ${\beta}_m^n, m \in \textbf{M}_n, n \in \textbf{N}$. ${\beta}_m^n = 1$ implies that subchannel $n$ is allocated to user $m$, and ${\beta}_m^n = 0$ means the other case. Ideally, more and more users are assigned to a subchannel, the better the system performance. However, due to the increasing level of interference with more and more assigned users and the decoding complication of the SIC technique, not necessarily more users assigned to a subchannel will enhance the system energy efficiency. While giving weight to this observation and insight, we assume that maximum $K$ users can be assigned to a subchannel\footnote{Setting $K$ to $M$, the multi-user diversity of a subchannel can be fully exploited. Due to the limited power level and the superposition-coded interference level caused by other users, an ideal resource allocation scheme should smartly select the users that are beneficial for that subchannel in terms of energy efficiency}. Therefore, if the achieved instantaneous rate of user $m$ is $R_m$ and its minimum required rate is $R_m^{\mbox{min}}$ (to satisfy its quality-of-service requirement), the energy-efficient downlink subchannel and power allocation problem in this context can be formulated as (\ref{eq:opt-prob1})-(\ref{eq:opt-prob5}). Note that the aggregate power level of subchannel $n$ is denoted by $p_n$, and $\sum_{m \in \textbf{M}_n}p_m^n = p_n$.

%

% \displaystyle\sum_{i \in \textbf{M}_n^m}p_i^ng_i^n}

\begin{eqnarray}
\label{eq:opt-prob1} &\displaystyle\max_{\{p_m^n, {\beta}_m^n\}}\displaystyle\sum_{n \in \textbf{N}}\frac{1}{p_c + \displaystyle\sum_{m \in \textbf{M}_n}p_m^n}\displaystyle\sum_{m \in \textbf{M}_n}{\beta}_m^n\mbox{log}\left(1 + \mbox{SINR}_m^n\right), \\
& \nonumber \mbox{subject to} \\
\label{eq:opt-prob2} & \displaystyle\sum_{m \in \textbf{M}}{\beta}_m^n \le K,~ \forall{n \in \textbf{N}},~~~\displaystyle\sum_{n \in \textbf{N}}{\beta}_m^n \le N,~ \forall{m \in \textbf{M}}, \\
\label{eq:opt-prob3} & {\beta}_m^n \in \{0, 1\},~ \forall{m \in \textbf{M}},~ \forall{n \in \textbf{N}}, \\
\label{eq:opt-prob4} & \displaystyle\sum_{n \in \textbf{N}}\displaystyle\sum_{m \in \textbf{M}_n}{\beta}_m^np_m^n \le p^{\mbox{max}},~ p_m^n \ge 0, \\
\label{eq:opt-prob5}& R_m  \ge R_m^{\mbox{min}}, \forall{m \in \textbf{M}}.
\end{eqnarray}

In the above formulation, there are two types of variables, i.e., $\{{\beta}_m^n\}$ and $\{p_m^n\}$. $\{{\beta}_m^n\}$ are the set of discrete variables, and the problem is NP-hard because of these variables. The NP-hardness property of this problem can be proved by mapping it to the classical relaxed bin-packing problem~\cite{SMartello1990}. Let us assume that the subchannels in set \textbf{N} are the possible bins, the users in set \textbf{M} are items and the capacity of each bin is $K$. Now, if we assume that an item can reside into multiple bins and if we transform the objective of the problem from minimizing the number of used bins to maximizing the system energy efficiency, reduction of our problem to the relaxed bin packing problem is completed. Thus, the NP-hardness property of our problem is proved. On the other hand, even if the information of set $\{{\beta}_m^n\}$ are known, it is straightforward to prove that the problem is non-convex with respect to (w.r.t.) $\{p_m^n\}$. Moreover, in order to reduce the complexity of the solution method proposed in the following section, the ordering of the power level and gain of the users in set $\textbf{M}_n$ are considered to be fixed and given as follows. For any $2$ users $m$ and $m'$ in set $\textbf{M}_n$, if they follow $g_m^n > g_{m'}^n$, they must hold $p_m^n < p_{m'}^n$.  

% because of the interference term $1 + \sum_{i \in \textbf{M}_n^m}p_i^ng_m^n$ inside the log term of (\ref{eq:opt-prob1})

%The solution of the aforementioned optimization problem is provided in the following section.

\section{Solution Approach}
\label{sec:sol}

In this section, we explore the solution approach of the energy-efficient downlink resource allocation problem of a NOMA system described in (\ref{eq:opt-prob1})-(\ref{eq:opt-prob5}). Apparently, due to the discrete nature of subchannel assignment (i.e., variables $\{{\beta}_m^n\}$) and the continuous nature of power assignment (i.e., variables $\{p_m^n\}$), this is an MINLP problem. Furthermore, the assignment issue of users on the same subchannel brings further complication in the solution strategy. Therefore, we realize that decomposing the problem into a subchannel allocation subproblem followed by a power loading subproblem and then solving each subproblem based on the insights of the optimal solution, is a good strategy to solve the joint problem. Consequently, based on the assumption that the maximal power of the BS ($p^{\mbox{max}}$) is equally subdivided among all the subchannels, we solve the subchannel-user mapping subproblem using a many-to-many matching model. Compared to the one-to-many matching model adopted in~\cite{FFang2016}, we believe that a many-to-many one can capture the structure of the problem well. This is because one user can be assigned to multiple subchannels in order to exploit the multi-user diversity of wireless systems, and one subchannel can be assigned to multiple users to take the advantages of the NOMA technology. While solving the subchannel-user mapping subproblem, given the power constraint of each subchannel, it is required to assign power optimally among the allocated users so that the overall rate of that subchannel is maximized. For this purpose, we have noticed that if the value of $K$ is larger than $2$, the resultant optimization problem cannot be written as the difference of convex functions, and hence the DC programming-based approach is not appropriate to solve this problem. However, we found that GP is a well-fit technique to solve this problem after approximating the objective function of the original problem by a ratio of two posynomials. For the solution of the second part of the problem, we assume that the subchannel-user mapping information are available. Given this information,~\cite{FFang2016} has utilized the DC programming-based approach to allocate power across the subchannels while ignoring the detailed granularity of the user power level. On the attempt of finding an elegant solution for the power loading subproblem, we find that the GP-based approach is a good one as well upon approximating the problem by a ratio of two posynomials. 

\subsection{Subchannel and User Mapping Scheme}
Intuitively, assignment of many subchannels to a user and allocating multiple users to a subchannel (to follow the guidelines of the NOMA technique) is expected to enhance the overall energy efficiency of the system. However, due to the usage of practical, finite-length and possibly non-Gaussian codes, we assume that maximum $K$ users can be assigned to a subchannel. Given the total power constraint of the BS, this problem is NP-hard. The nature of the problem implies that a many-to-many matching model~\cite{KHamidouche2014, ARoth1984} is appropriate to capture the aforementioned behavior. Given that maximum $K$ users can be multiplexed to a subchannel, $M$ users in set \textbf{M} and $N$ subchannels in set \textbf{N} are two sets of players of this many-to-many matching model. Note that each user $m$ can have infinite ($N$ in practice) subchannels if possible. In this case, since the BS has the maximal power constraint $p^{max}$, this should be subdivided equally among all the subchannels. 

\noindent
\textbf{Definition 1:} A many-to-many matching model $\mu$ is a mapping from set \textbf{M} to set \textbf{N} such that every $m \in \textbf{M}$ and $n \in \textbf{N}$ satisfy the following properties:

\begin{itemize}

\item ${\mu}(m) \subseteq \textbf{N}$ and ${\mu}(n) \subseteq \textbf{M}$

\item $|{\mu}(m)| \le \infty, \forall{m \in \textbf{M}}$

\item $|{\mu}(n)| \le K, \forall{n \in \textbf{N}}$

\item $n \in {\mu}(m)$ if and only if $m \in {\mu}(n)$

\end{itemize}

\noindent
where ${\mu}(m)$ is the set of partners for user $m$ and ${\mu}(n)$ is the set of partners for subchannel $n$ under the matching model $\mu$. The definition states that each user in set \textbf{M} is matched to a subset of subchannels in set \textbf{N}, and vice versa. However, before accomplishing these assignment operations, each user needs to have a preference list based on some criteria. The criterion of constructing the preference list for the users is based on their gain from the subchannels, which is similar to that in~\cite{FFang2016}. The preference of each subchannel $n$ is based on the overall benefit (i.e., energy efficiency) of the system. For example, if user $m$ chooses subchannel $n$, this subchannel only accepts this user if and only the energy efficiency of the system is enhanced by this allocation. 

% In other words, each user may choose a set of subchannels as communication media, whereas each subchannel may choose a set of users to be assigned with in order to maximize the overall energy efficiency of the system.

% We use the notation ${{\bm{\Omega}}'}_m {\succ} {{\bm{\Omega}}''}_m$ to imply that user $m$ wants to have the subchannels in subset ${{\bm{\Omega}}'}_m$ than the subchannels in subset ${{\bm{\Omega}}''}_m$, where ${{\bm{\Omega}}'}_m \subseteq \textbf{N}$ and ${{\bm{\Omega}}''}_m \subseteq \textbf{N}$. Similar analogy can be made for any subchannel $n$ in set \textbf{N}.

To solve our subchannel-user mapping problem, we are interested to look at a stable solution, in which there are no players that are not matched to one another but they all prefer to be partners. Since the subchannel players give preference to the overall energy efficiency of the system while choosing partners from set \textbf{M}, a stable solution is considered to be an elegant solution of this problem\footnote{At this point, none of the players can enhance their performance further by choosing alternative partners, and hence this point is considered as a stable point.}. In the many-to-many matching models~\cite{ARoth1984}, many stability concepts can be considered based on the number of players that can improve their utility by forming new partners among one another. However, due to the large number of players ($\textbf{M} \cup \textbf{N}$) in our problem, identifying optimal subset of partners for a player is computationally intractable. Consequently, we choose to solve the matching problem by identifying the partners one by one from the opposite set. This way of choosing the partners in the matching relation brings pair-wise stability. In \textit{Definition 1} and \textit{Definition 2}, we highlight some properties of a pair-wise stable matching relation. For the sake of these definitions, we define some notations. For example, $\textbf{C}_m(\textbf{N})$ denotes the choice set of user $m$, which basically follows $\textbf{C}_m(\textbf{N}) \subseteq \textbf{N}$. Moreover, ${\textbf{M}'}_n {\succ}_m {\textbf{M}''}_n$ implies that subchannel $n$ prefers the users in set  ${\textbf{M}'}_n$ over that in set ${\textbf{M}''}_n$.

\noindent
\textbf{Definition 1:} Consider that pair $(m,n)$ is not the element of matching model $\mu$, that is $m \not\in {\mu}(n)$ and $n \not\in {\mu}(m)$. Now consider another pair $(\phi, \varphi)$, that satisfies $\phi \in \textbf{C}_m({\mu}(n) \cup \{n\})$ and $\varphi \in \textbf{C}_n({\mu}(n) \cup \{m\})$. For the matching relation $\mu$ to be pairwise stable, it is not possible that both $\phi {\succ}_m {\mu}(m)$ and $\varphi {\succ}_n {\mu}(n)$ are satisfied.

\noindent
\textbf{Definition 2:} Let ${\textbf{M}}_n$ (${\textbf{M}}_n \in \textbf{M}$) be the set of matched users of subchannel $n$ under the matching relation $\mu$, and $|{\textbf{M}}_n| = K$. Consider $m$ is one of the matched users in set ${\textbf{M}}_n$. Now, another user $m' \in \textbf{M}$ has come to be matched with subchannel $n$. Note that $m' \not\in {\textbf{M}}_n$. However, if user $m$ is replaced by user $m'$, the performance of subchannel $n$ is enhanced. Therefore, in order to have stable matching, ${\textbf{M}}_n = ({\textbf{M}}_n /\ m) \cup m'$ should hold. This phenomenon of stable matching is called substitutability.

% Note that in this algorithm, we are interested in pair-wise stability, and hence the players (i.e., users and subchannels) choose their partners one-by-one instead of a subset.

% Over the iterations, these sets are filled by subchannel-user assignment information.

While satisfying the properties of a stable many-to-many matching relation (e.g, substitutability~\cite{ARoth1984}), we have proposed an algorithm in \textit{Algorithm~\ref{alg:sc-usr-map}}. In order to bring the stability in this matching relation or enhance the overall energy efficiency of the system, we have adopted a few heuristics or strategies. The description of the algorithm is as follows. In order to realize the outcome of each step of the algorithm, we introduce one more type of set variable: ${\bm{\Omega}}_m$ that holds the allocated subchannels for user $m$. First, ${\bm{\Omega}}_m, m \in \textbf{M}$ and $\textbf{M}_n, n \in \textbf{N}$ are initialized with $\emptyset$. These sets are gradually filled up as we go through the iterations in between step $3$ and step $26$. At the initialization phase, each user $m \in \textbf{M}$ also constructs its subchannel preference list based on the descending order of their gain. Then, inside the outer-most loop (between step $3$ and step $26$), if no assignment is possible, the algorithm terminates\footnote{At this point, it is assumed that the system has reached a stable situation or the improvement of energy efficiency is no longer possible.}. Inside the inner loop (between step $4$ and step $25$), each user $m$ chooses its most preferred unallocated (and not rejected already) subchannel $n$. At this point, two conditions are possible. The first condition is that the number of allocated users to subchannel $n$ can be less than $K$ (maximum allowable number of users per subchannel), and the second condition is the other case. If the first condition is true, we can apply the addition strategy for this subchannel-user assignment: user $m$ can be added to this subchannel and this strategy is added to set \textbf{S} (which was initialized before sorting out the possible strategies for use $m$). Whereas, for the second condition, only substitution strategy (already allocated user $m'$ can be replaced by new user $m$) is possible. After filling the strategy set \textbf{S}, no matter the number of allocated users to subchannel $n$ is less than or equal to $K$, the elements of \textbf{S} are filtered based on the previous sum-rate of subchannel $n$ before this new possible assignment. The filtered strategy set is \textbf{CS} in \textit{Algorithm~\ref{alg:sc-usr-map}}. Finally, $s^{\mbox{Best}}$ strategy is chosen based on the sum-rate (${\Gamma}_{s}^n$ in \textit{Algorithm~\ref{alg:sc-usr-map}}) each strategy achieves. If $s^{\mbox{Best}}$ is empty, the innermost loop continues, and the next user is chosen from set \textbf{M} for building its possible strategy set. For the other case, the corresponding strategy is executed. As a result, set ${\bm{\Omega}}_m$, set ${\bm{\Omega}}_{m'}$ (only for the substitution strategy) and set $\textbf{M}_n$ are updated. By analyzing the algorithm, we conclude \textit{Proposition 1}, \textit{Proposition 2} and \textit{Theorem 1}.

%  using many-to-many matching model

\begin{algorithm}
\caption{The energy-efficient subchannel-user mapping algorithm via a many-to-many matching model.}
\label{alg:sc-usr-map}
\begin{algorithmic}[1]
\STATE ${\bm{\Omega}}_m \leftarrow \emptyset, \forall{m \in \textbf{M}}$; $\textbf{M}_n \leftarrow \emptyset, \forall{n \in \textbf{N}}$.
\STATE Each user $m \in \textbf{M}$ produces its preference list in the descending order of its gain on subchannel $n \in \textbf{N}$
%Combining the preference list of all users in \textbf{M}, preference matrix $\bm{\Lambda}$ is constructed.

\REPEAT
%\FOR{$i \leftarrow 1~\mbox{to}~N$}
\FOR{$m \in \textbf{M}$}

\STATE $n$ $\leftarrow$ The best unallocated (and not rejected previously) subchannel of user $m$. 
% \COMMENT{\textbf{S} is the strategy set}
% \COMMENT{\textbf{CS} is the candidate strategy set}
\STATE ${\Gamma}_{\mbox{prev}}^n \leftarrow$ The aggregate rate of subchannel $n$.
\STATE $\textbf{S} \leftarrow \emptyset$, $\textbf{CS} \leftarrow \emptyset$, and $\textbf{TRate} \leftarrow \emptyset$. 
% \COMMENT{\textbf{Thrput} is the set which contains the sum-rate incurred by each strategy in set \textbf{CS}}
\IF{The number of assigned users on subchannel $n$ is less than $K$}
\STATE Construct strategy $s$ by adding user $m$ to set $\textbf{M}_n$, and add $s$ to set \textbf{S}.
\ELSIF{The number of assigned users on subchannel $n$ is equal to $K$}
\STATE Construct each strategy $s$ by replacing each user $m' \in \textbf{M}_n$ by user $m$, and add $s$ to set \textbf{S}.
\ENDIF
\FOR{$s \in \textbf{S}$}
\STATE Due to strategy $s$, determine the power level of user $m$ in $\textbf{M}_n$ using (18)~\cite{FFang2016} or the approach in Section~\ref{sssec:GP-indiv}, and ${\Gamma}_{s}^n \leftarrow$ The aggregate rate of subchannel $n$ due to strategy $s$.
\STATE Add ${\Gamma}_{s}^n$ to set \textbf{TRate} and add strategy $s$ to \textbf{CS} if and only if ${\Gamma}_{s}^n - {\Gamma}_{\mbox{prev}}^n > 0$.
% $rate$ is larger than the sum-rate of subchannel $n$ before applying strategy $s$. 
\ENDFOR

\STATE $s^{\mbox{Best}} \leftarrow \argmax_{s \in \textbf{CS}}\textbf{TRate}(s)$.
\IF{$s^{\mbox{Best}}$ is the substitution strategy}
\STATE ${\bm{\Omega}}_m \leftarrow {\bm{\Omega}}_m \cup \{n\}$, ${\bm{\Omega}}_{m'} \leftarrow {\bm{\Omega}}_{m'} /\ \{n\}$, and $\textbf{M}_n \leftarrow (\textbf{M}_n /\ \{m'\}) \cup \{m\}$. \COMMENT{$m'$ is the to-be-replaced user and $m$ is the new user on subchannel $n$}
\STATE Update the preference list of user $m$ and user $m'$.
%, that indicates the allocation (or rejection) of subchannel $n$ in matrix $\bm{\Lambda}$.  
\ELSIF{$s^{\mbox{Best}}$ is the addition strategy}
\STATE ${\bm{\Omega}}_m \leftarrow {\bm{\Omega}}_m \cup \{n\}, \textbf{M}_n \leftarrow \textbf{M}_n \cup \{m\}$.
\STATE Update the preference list of user $m$.
%, that indicates the allocation of subchannel $n$ in matrix $\bm{\Lambda}$.
\ENDIF
%\ENDFOR
\ENDFOR
\UNTIL{The overall performance of the system cannot be enhanced}
\end{algorithmic}
\end{algorithm}

% All users in set \textbf{M} approach all subchannels in set \textbf{N}

%\begin{figure*}%
%\centering
%\begin{subfigure}{0.6\columnwidth}
%\includegraphics[width=\columnwidth]{figures/prop1_proof.eps}%
%\caption{A sample example that $[R_n(\{1,2,3\})-R_n(\{1,2\})]$ can be both positive and negative.}%
%\label{fig:prop1-proof}%
%\end{subfigure}\hfill%
%\begin{subfigure}{0.6\columnwidth}
%\includegraphics[width=\columnwidth]{figures/prop2_proof.eps}%
%\caption{A sample example that $R_n(\{1,2\})$ is increasing w.r.t. $g_2^n$.}%
%\label{fig:prop2-proof}%
%\end{subfigure}\hfill%
%\begin{subfigure}{0.6\columnwidth}
%\includegraphics[width=\columnwidth]{figures/log_expansion.eps}%
%\caption{A sample verification of natural log series~\cite{RNave}.}%
%\label{fig:log-expansion}%
%\end{subfigure}%
%\caption{}
%\label{figabc}
%\end{figure*}

%\begin{figure*}
%  \centering
%    \subfigure[a\label{fig:prop1-proof}]{\includegraphics[scale=0.38]{figures/prop1_proof.eps}}\quad
%    \subfigure[a\label{fig:prop2-proof}]{\includegraphics[scale=0.38]{figures/prop2_proof.eps}}\quad
%    \subfigure[c\label{fig:log-expansion}]{\includegraphics[scale=0.38]{figures/log_expansion.eps}}
% \caption{t}
%  %\label{main figure label}
%\end{figure*}

\noindent
\textbf{Proposition 1:} Given a certain value of $K$, allocating more and more users to a subchannel cannot enhance the energy efficiency further.

% although the allowable number of users per subchannel (i.e., $K$) is large

\noindent
\textbf{Proof:} See \textit{Appendix A}.

% Note that in step $4$, we always choose the best unallocated subchannel-user tuple in terms of their gain. Therefore, it is impossible that $g_3^n > g_1^n$ or $g_2^n > g_1^n$ is true. The maximum value of $g_3^n$ could be that of $g_2^n$.

\noindent
\textbf{Proposition 2:} Once a user is rejected while to be assigned with a subchannel through the addition or the substitution strategy, that rejection is final. More precisely, even if that user wants to be assigned to that particular subchannel later, this assignment no longer enhances the system energy efficiency.

\noindent
\textbf{Proof:} See \textit{Appendix B}.

\noindent
\textbf{Theorem 1:} The subchannel-user mapping algorithm (i.e., \textit{Algorithm 1}) is guaranteed to converge to a pair-wise stable matching relation.

\noindent
\textbf{Proof:} See \textit{Appendix C}.

\noindent
\textbf{\textit{Computational Complexity of \textit{Algorithm~\ref{alg:sc-usr-map}}:}} As discussed and proved in \textit{Proposition 2}, if user $m$ is rejected by subchannel $n$ once, that user does not propose subchannel $n$ to be matched with anymore. This is because the energy efficiency of this subchannel $n$ is not enhanced in any way if previously rejected user is given preference in the following iterations. The enhancement of the energy efficiency for each subchannel implies the enhancement of the overall energy efficiency in the system. The termination of the outer-most loop (between step $3$ and step $26$) depends on the improvement of the overall energy efficiency. To be precise, if user $m$ in the inner loop proposes a subchannel $n$ and the corresponding $s^{\mbox{Best}}$ results in empty value, the next user in set \textbf{M} builds its possible strategy set. If none of the users in set \textbf{M} can be matched with any of the subchannel in set \textbf{N}, the outer-most loop terminates. Therefore, intuitively, each user $m \in \textbf{M}$ approaches at most $N$ subchannels to be matched with. Consequently, the outer-most loop runs at most $MN$ times. On the other hand, the complexity of finding the best strategy for each user inside the inner loop is $\mbox{O}(K)$, and the corresponding reason is as follows. If the number of users assigned to subchannel $n$ is less than $K$, only one strategy (the addition strategy) is possible. If the number of users assigned to subchannel $n$ is exactly $K$, then $K$ strategies are possible due to the $K$ possible replacement policies. Since the value of $K$ is usually not large in practice, we consider the complexity of finding the best strategy as constant. The remaining operations inside the inner-most loop happen in constant time, and so we can ignore the complexity of these operations as well. Consequently, the overall complexity of the algorithm is $\mbox{O}(MN)$.

\subsection{Power Allocation Scheme}
\label{ssec:GP}

From \textit{Algorithm~\ref{alg:sc-usr-map}}, we know the subchannel-user mapping information, i.e., $\textbf{M}_n, n \in \textbf{N}$ and ${\bm{\Omega}}_m, m \in \textbf{M}$. This information is derived based on the assumption that the maximal power level $p^{\mbox{max}}$ of the BS is equally subdivided among all subchannels, i.e., $p_n = p^{\mbox{max}}/N, n \in \textbf{N}$. However, in (\ref{eq:opt-prob1}), we see that the instantaneous rate of user $m$ assigned to subchannel $n$ is the increasing function of $p_m^n$ and the decreasing function of the interference power level caused by other users, i.e., $p_i^n, i \in \textbf{M}_n^m$. Consequently, even if the information about $\textbf{M}_n, n \in \textbf{N}$ and ${\bm{\Omega}}_m, m \in \textbf{M}$ are known, the power allocation across all subchannel-user slots, i.e., $p_m^n, m \in \textbf{M}, n \in \textbf{N}$, is a non-convex optimization problem. In the literature, the transformation of such a type of problem to convex form is not reported yet, and hence the dual-based method~\cite{CWang2016} or the bisection search~\cite{QSun22015} cannot be adopted to solve this problem. Therefore, our next objective is to design a method that allocates power across all the subchannel-user slots in such a manner that near-optimality is achieved. We have adopted the GP-based optimization technique to solve this power allocation problem, the description of which is provided in the following discussions.

\begin{figure}
  \begin{center}
    \includegraphics[width=0.5\columnwidth]{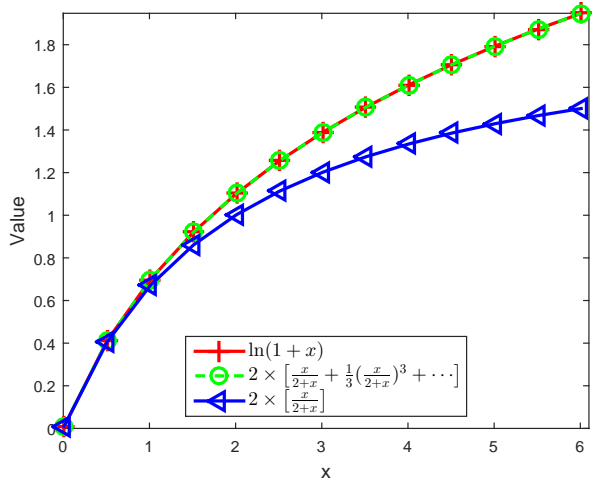}
    \caption{A sample verification of natural log series expansion~\cite{RNave}.}
    \label{fig:log-expansion}
  \end{center}
\end{figure}

% The name of the functions, with which GP deals, are monomials and posynomials. 

A GP~\cite{Boyd2007, MChiang2005} is a type of mathematical optimization problem with the characteristics that the objective and constraint functions must be monomial(s) or posynomial(s). Two problems that we have discussed in the following have a certain form, such as $\frac{G_n(q_1,\cdots,q_T)}{G_d(q_1,\cdots,q_T)}$. Here, both functions $G_n(q_1,\cdots,q_T)$ and $G_d(q_1,\cdots,q_T)$ are posynomials; $q_1,\cdots,q_T$ are the variables in the problem; and $T$ is the number of variables. This type of function is still not amenable to GP. To make this problem amenable to GP, there are some heuristics, such as the single condensation method and the double condensation method~\cite{Boyd2007, MChiang2005}. We plan to apply the single condensation method in this context, which requires the denominator (i.e., $G_d(q_1,\cdots,q_T)$) to be approximated via some monomial ${\lambda}\prod_{i=1}^T{q_i}^{a_i}$. Given the values of $q_1,\cdots,q_T$, the unknown variables of this monomial can be approximated by $a_i = \frac{q_i}{G_d(.)}\frac{{\partial}G_d(.)}{{\partial}q_i}$ and $\lambda = \frac{G_d(.)}{\prod_{i=1}^T{q_i}^{a_i}}$. As a result, the approximated objective function becomes the ratio of a posynomial and a monomial, which is a posynomial. Consequently, we require an iterative process in order to solve this problem step by step. The steps of this iterative process are provided in Section III-A of~\cite{RRuby20152}.

\begin{comment}

\begin{enumerate}
\item Set the initial values for variables $q_1,\cdots,q_T$.
\item Determine the values of $a_1,\cdots,a_T$, and $\lambda$.
\item Solve the problem associated with resultant objective function, i.e., the ratio of the posynomial and the monomial, via GP.
\item If the absolute difference between the current and previous values of the variables is less than or equal to some small number, i.e., $\epsilon$, go to step $2$, and if otherwise, stop the iterative process. 
\end{enumerate}

The final values of the variables obtained in the last iteration of the iterative process is the solution of our defined optimization problem.
\end{comment}

\subsubsection{Joint Power Allocation among all Subchannels}
\label{sssec:GP-joint}

Once the subchannel-user assignment information, i.e., $\textbf{M}_n, n \in \textbf{M}_n$, in order to obtain the energy-efficient power level of each  subchannel-user tuple given the power constraint at the BS, we need to solve an optimization problem, which is as follows.

\begin{eqnarray}
\label{eq:GP-joint}
\nonumber & \displaystyle\max_{\{p_m^n\}}\displaystyle\sum_{n \in \textbf{N}}\frac{1}{p_c+\displaystyle\sum_{m \in \textbf{M}_n}p_m^n}\displaystyle\sum_{m \in \textbf{M}_n}\mbox{log}_2\left(1 + \frac{g_m^np_m^n}{{\sigma}_n^2 + \displaystyle\sum_{i\in\textbf{M}_n^m}p_i^ng_m^n}\right),  \\
\nonumber & \mbox{s.t.}~\displaystyle\sum_{n \in \textbf{N}}\displaystyle\sum_{m \in \textbf{M}_n}p_m^n \le p^{max}, \\
& \displaystyle\sum_{m\in \textbf{M}_n}p_m^n \ge p_n^{\mbox{min}},~\forall{n \in \textbf{N}}.
\end{eqnarray}

\begin{equation}
\mbox{log}_2\left(1 + \frac{g_m^np_m^n}{{\sigma}_n^2 + \displaystyle\sum_{i\in\textbf{M}_n^m}p_i^ng_m^n}\right) = \frac{\frac{g_m^np_m^n}{{\sigma}_n^2 + \displaystyle\sum_{i\in\textbf{M}_n^m}p_i^ng_m^n}}{2 + \frac{g_m^np_m^n}{{\sigma}_n^2 + \displaystyle\sum_{i\in\textbf{M}_n^m}p_i^ng_m^n}}
\end{equation}

% \displaystyle\sum_{m \in \textbf{M}_n}
%
%\begin{figure}
%  \begin{center}
%    \includegraphics[width=0.8\columnwidth]{log_expansion.eps}
%    \caption{A sample verification of natural log series~\cite{RNave}.}
%    \label{fig:log-expansion}
%  \end{center}
%\end{figure}

%Note that the last constraints in the aforementioned optimization problem are the transformed version of the constraints in (\ref{eq:opt-prob5}). For this, we require to solve $M$ equations in (\ref{eq:opt-prob5}) jointly.

The relations in~(\ref{eq:opt-prob5}) basically are associated with $M$ equations. The number of variables (i.e., $\{p_m^n\}_{m \in \textbf{M}_n, n \in \textbf{N}}$) in all these equations is $\sum_{n \in \textbf{N}}|\textbf{M}_n|$. However, using (18)~\cite{FFang2016}, the number of variables in all these equations can be reduced to $N$. Therefore, for $M \ge N$, intuitively, by solving all these equations jointly, the relations in the last constraint of~(\ref{eq:GP-joint}) are achievable. For the other case (i.e., $M < N$), the last constraint of~(\ref{eq:GP-joint}) can be approximated as well based on some simple assumption. Moreover, on subchannel $n$, the set of users, $\textbf{M}_n^m$, that cause interference to user $m$ is determined based on the decoding order described in Section~\ref{sec:sysmodel}. Clearly, the objective function here is not amenable to GP (not even any GP-based heuristic method) because of the energy efficiency factor $\frac{1}{p_c+\sum_{m \in \textbf{M}_n}p_m^n}$. Therefore, we have decided to expand the log series. Since the SINR of user $m$ on subchannel $n$ is larger than zero, we can expand the log function according to $\mbox{ln}(1+x) = 2\left[\frac{x}{2+x} + \frac{1}{3}(\frac{x}{2+x})^3+\cdots\right]$~\cite{RNave}. The accuracy of the log expansion is verified in Fig.~\ref{fig:log-expansion}. For the sake of simplicity, using the first term of the expanded log function, we transform the original objective function to a form, which becomes the ratio of two posynomials. As shown in Fig.~\ref{fig:log-expansion}, both the original log function and the first term of the expanded series are increasing with the increasing value of $x$. Moreover, the difference between the original log function and the first term of the log series increases with the increasing value of $x$. However, this difference is negligible when the value of $x$ is $\le 3$. If we look at the objective function in ~(\ref{eq:GP-joint}), $x$ is mapped to the SINR term of each user on a particular subchannel. The SINR of a user on a subchannel is a function of its and other users' gain and assigned power level. If the values of gain and maximal power level of the system are such that the SINR of a user on a subchannel does not exceed $5$ dB (the absolute value of which is around 3), the resultant approximated problem using the first term of the log series does not deviate much from the original problem. On the other hand, the search range of the SINR for a user over a subchannel may exceed $5$ dB at higher gains and a larger power level of the system. However, the first term of the log series dominates the entire log function. Moreover, since the nature of the log function as well as the first term of the log series are increasing, the sum of the log function as well as the sum of the first term (of the log series) are increasing. Therefore, when the search range of a SINR exceeds $5$ dB, although the approximated function is deviated from the original function, the resultant solution of the approximated problem should not be very bad. The accuracy of the resultant solution can be improved by approximating the problem using the first two terms of the log series instead of only the first one. Since the problem is a maximization problem, we apply the inverse function\footnote{The inverse function of $\frac{A}{B}$ is $\frac{B}{A}$.} to transform it to the minimization one. Consequently, the problem becomes solvable via GP, as described in Section~\ref{ssec:GP}. The outcome of the proposed fine-grained power allocation scheme strengthens the well-established truth~\cite{SHan2014} that the two objectives of maximizing the energy efficiency and maximizing the spectrum efficiency cannot be achieved simultaneously at the same operating point.

%In \textit{Proposition 3}, we provide one property of the optimal fine-grained power allocation concept in this context.

%  Typically, in practice, the SINR of a particular user on a subchannel is at most $5$ dB, the absolute value of which is around $3$. Therefore, although the objective function is the sum of log functions, at $\le 5$ dB SINR

% at the SINR values, that are greater than $5$ dB, the approximation may be incorrect.

% On the other hand, our objective is to find the optimal value of $x$, in which the objective function is maximized. Therefore, intuitively, at the same value of $x$, both the original objective function and the transformed objective function using the first term of log series should incur the optimal value. As a result, because of taking the first term of log series, the correctness of the transformed problem is not deviated from the original problem.

%the two objectives of maximizing energy efficiency and maximizing spectrum efficiency are not achievable simultaneously

\begin{comment}

\noindent
\textbf{Proposition 3:} If the overall energy efficiency achieved by a certain power allocation scheme is better than that of some benchmark scheme, it does not necessarily implies that the overall throughput achieved by that scheme will be better compared to the benchmark scheme.

\noindent
\textbf{Proof:} See \textit{Appendix D}.
\end{comment}

\subsubsection{Power Allocation of a Subchannel}
\label{sssec:GP-indiv}

% \displaystyle\max_{\{p_m^n\}}\displaystyle\sum_{m \in \textbf{M}_n}\mbox{log}_2\left(1 + \frac{g_m^np_m^n}{1 + \sum_{i\in\textbf{M}_n^m}p_i^ng_m^n}\right) =

For subchannel $n$, once $\textbf{M}_n$ and $p_n$ are known, we would like to determine $\{p_m^n\}_{m \in {\textbf{M}}_n}$. If $K=2$,~\cite{FFang2016} has proved that this problem can be solved using the DC programming-based approach. However, if $K > 2$, this problem can no longer be solvable via this approach. To prove this statement, let us assume $K=3$, and $g_1^n > g_2^n > g_3^n$ holds for subchannel $n$. Under this circumstances, the instantaneous rate of the $3$rd user on subchannel $n$ can be given by $\mbox{log}_2(1 + g_3^np_3^n/(1+g_3^np_1^n+g_3^np_2^n))$. The interference term in the instantaneous rate of the $3$rd user is a barrier to express the problem in terms of a difference of convex functions. If the value of $K$ is even larger, the nonconforming nature of the problem to the DC programming-based approach becomes even more severe. In general, the energy-efficient power allocation problem of subchannel $n$ can be written as follows.

% after some transformation based on the fact that $\mbox{log}_2(A) + \mbox{log}_2(B) = \mbox{log}_2(AB)$,

\begin{eqnarray}
\label{eq:GP-indiv1}&  \displaystyle\max_{\{p_m^n\}}\frac{1}{p_c+\displaystyle\sum_{m \in \textbf{M}_n}p_m^n}\displaystyle\sum_{m \in \textbf{M}_n}\mbox{log}_2\left(1 + \frac{g_m^np_m^n}{{\sigma}_n^2 + \sum_{i\in\textbf{M}_n^m}p_i^ng_m^n}\right), \\
\label{eq:GP-indiv2}&\mbox{s.t.}~\displaystyle\sum_{m \in \textbf{M}_n}p_m^n \le p_n.
\end{eqnarray}

Expanding the log function and then taking the first term of the series, the objective function in (\ref{eq:GP-indiv1})-(\ref{eq:GP-indiv2}) becomes a ratio of two posynomials. Upon applying the inverse function, the objective function is transformed to a form (i.e., $\frac{G_n(.)}{G_d(.)}$) as described in Section~\ref{ssec:GP}. 

% In order to transform the maximization problem in~(\ref{eq:GP-indiv1})-(\ref{eq:GP-indiv2}) to the minimization one, we just need to apply inverse function on the problem. After this

\subsubsection{Joint Computationally-Efficient Suboptimal Power Allocation among all Subchannels}
\label{sssec:sopt-joint}

The solution of the fine-grained power allocation problem in Section~\ref{sssec:GP-joint} is computationally intensive due to the cumbersome iterative process of the single condensation method as well as the implementation complication of the off-the-shelf GP solvers~\cite{cvx}. The complexity of this iterative process is proportional to the number of degrees of freedom in the system. If the number of users and subchannels in the system is moderately large, the solution of the problem cannot be obtained in a reasonable amount of time. Consequently, in this subsection, we have proposed another suboptimal solution for the fine-grained power allocation across the subchannels and users with much less computational overhead even with the off-the-shelf GP solvers~\cite{cvx}. The mechanism of this suboptimal solution is developed based on the insights of the optimal solution, the description of which is given as follows. The idea of this suboptimal solution is to obtain the energy-efficient solution of the subchannels one by one. For this, using (18)~\cite{FFang2016}, we plan to approximate the problem in (\ref{eq:GP-indiv1})-(\ref{eq:GP-indiv2}) to a one-variable optimization problem no matter the value of $|\textbf{M}_n|$ is. If the energy-efficient power level of subchannel $n$ is $p_n$, using (18)~\cite{FFang2016}, we have $p_m^n = {\gamma}_m^np_n,~m \in \textbf{M}_n$, where $\sum_{m\in\textbf{M}_n}{\gamma}_m^n = 1$. Then, we can transform the problem in (\ref{eq:GP-indiv1})-(\ref{eq:GP-indiv2}) to either the problem in (\ref{eq:sopt-DC1})-(\ref{eq:sopt-DC2}) or that in (\ref{eq:sopt-GP1})-(\ref{eq:sopt-GP2}). Since we do not know the allocated power level of subchannel $n$ in advance, we assume that the maximal power level at the BS, $p^{\mbox{max}}$, is the upper limit to be allocated at the beginning. It is proved in~\cite{FFang2016} that $\mbox{log}_2\left(1 + p_nA_m(\{{\gamma}_m^n\},\{g_m^n\},{\sigma}^2) \right)/(p_c+\sum_{m \in \textbf{M}_n}{\gamma}_m^np_n)$ is a concave function, and hence the DC-programming-based approach can be used to solve the problem in (\ref{eq:sopt-DC1})-(\ref{eq:sopt-DC2}). On the other hand, in (\ref{eq:sopt-GP1})-(\ref{eq:sopt-GP2}), both the $C(\{{\gamma}_m^n\},\{g_m^n\},p_n,{\sigma}^2)$ and $D(\{{\gamma}_m^n\},\{g_m^n\},p_n,{\sigma}^2)$ are the polynomial functions of variable $p_n$. Using the first term of the log series, this problem can be transformed to a form which is the ratio of two posynomials as well. Consequently, we can adopt the GP-based single condensation method to obtain an elegant energy efficiency of subchannel $n$. From our experiments, the latter solution approach provides better solution for this problem, and so we adopt it as a part of our proposed suboptimal energy-efficient power allocation scheme. The number of iterations required for the single condensation method is constant as the number of optimization variable is one in this case no matter the value of $|\textbf{M}_n|$ for subchannel $n$ is. 

\begin{eqnarray}
\label{eq:sopt-DC1}&  \displaystyle\max_{p_n}\frac{1}{p_c+\displaystyle\sum_{m \in \textbf{M}_n}{\gamma}_m^np_n}\displaystyle\sum_{m\in\textbf{M}_n}\frac{\mbox{log}_2\left(1 + p_nA_m(\{{\gamma}_m^n\},\{g_m^n\},{\sigma}^2) \right)}{\mbox{log}_2\left(1 + p_nB_m(\{{\gamma}_m^n\},\{g_m^n\},{\sigma}^2)\right)}, \\
\label{eq:sopt-DC2}&\mbox{s.t.}~\displaystyle\sum_{m \in \textbf{M}_n}{\gamma}_m^np_n \le p^{\mbox{max}}.
\end{eqnarray}

\begin{eqnarray}
\label{eq:sopt-GP1} &  \displaystyle\max_{p_n}\frac{1}{p_c+\displaystyle\sum_{m \in \textbf{M}_n}{\gamma}_m^np_n}\mbox{log}_2\left(1 + \frac{C(\{{\gamma}_m^n\},\{g_m^n\},p_n,{\sigma}^2)}{D(\{{\gamma}_m^n\},\{g_m^n\},p_n,{\sigma}^2)}\right), \\
\label{eq:sopt-GP2}&\mbox{s.t.}~\displaystyle\sum_{m \in \textbf{M}_n}{\gamma}_m^np_n \le p^{\mbox{max}}.
\end{eqnarray}

\begin{comment}

convex nature of the energy efficiency problem of a subchannel. Given the power constraint of $p^{\mbox{max}}$, the energy-efficient power allocation of subchannel $n$ can be obtained by solving the optimization problem in~(\ref{eq:sopt-indiv}). 
\begin{eqnarray}
\label{eq:sopt-indiv}
\nonumber &  \displaystyle\max_{\{p_m^n\}}\displaystyle\sum_{m \in \textbf{M}_n}\mbox{log}_2\left(1 + \frac{g_m^np_m^n}{{\sigma}_n^2 + \sum_{i\in\textbf{M}_n^m}p_i^ng_m^n}\right), \\
&\displaystyle\sum_{m \in \textbf{M}_n}p_m^n \le p^{\mbox{max}}.
\end{eqnarray}
It is proved in~\cite{FFang2016} that the problem
in~(\ref{eq:sopt-indiv}) is convex, and hence it is straightforward to find the solution of this problem.

\end{comment}

After solving the problem in (\ref{eq:sopt-GP1})-(\ref{eq:sopt-GP2}) for all the subchannels, the resultant total power $\sum_{n \in \textbf{N}}\sum_{m \in {\textbf{M}}_n}{\gamma}_m^np_n$ can be $>$ or $\le$ $p^{\mbox{max}}$. For the latter case, the resultant total power, $\{{\gamma}_m^np_n\}_{n \in {\textbf{M}}_n, n \in \textbf{N}}$, can be considered as an elegant solution. On the other hand, for the former case, we need to revisit the solution of the optimization problem further. In order to develop a good solution for this case, we need to know the energy efficiency rate per unit of power level for all subchannels. Once these information are known, we can develop a suboptimal scheme, that assigns unit power to the subchannels one by one until the total allocated power to all the subchannels results in $p^{\mbox{max}}$. In order to know which subchannel has higher rate of energy efficiency, we construct an energy efficiency rate matrix \textbf{EEM}, the dimension of which is $(\lfloor{p^{\mbox{max}}/{\Delta}}\rfloor+1) \times N$. The $1$st row of this \textbf{EEM} matrix contains the energy efficiency of the subchannels given that each subchannel is assigned with $\Delta$ level of power. Each element of the $2$nd row contains the subtracted energy
efficiency when the corresponding subchannel is allocated with
$2\Delta$ and $\Delta$ level of power. Finally, each element of the last row contains the subtracted energy efficiency of the
corresponding subchannel when it is assigned to $\lfloor{p^{\mbox{max}}/{\Delta}}\rfloor \Delta$ and $p^{\mbox{max}}$ level of power. The energy efficiency of each subchannel is computed using the formulation in (\ref{eq:sopt-GP1})-(\ref{eq:sopt-GP2}) given the maximal power constraint designated for each element in matrix \textbf{EEM}. Based on the information of this matrix, power is assigned to each subchannel one by one in an iterative manner. For example, in one iteration, $(a, n),~a \in [1,
\lfloor{p^{\mbox{max}}/{\Delta}}\rfloor+1]$ tuple in \textbf{EEM} matrix has been selected since it has the highest value. In this case, subchannel $n$ is assigned to $a\Delta$ level of power. This process continues until the allocated power level to all subchannels, $\sum_{n \in \textbf{N}}p_n$, is equal to $p^{\mbox{max}}$. The detailed  steps of the entire iterative process are provided in \textit{Algorithm \ref{alg:sopt-pwr-alloc}}.

\begin{algorithm}[h!]
\caption{The computationally-efficient fine-grained energy-efficient power allocation scheme.}
\label{alg:sopt-pwr-alloc}
\begin{algorithmic}[1]
\FOR{$n \in \textbf{N}$}
\STATE Solve the problem in~(\ref{eq:sopt-GP1})-(\ref{eq:sopt-GP2}) for subchannel $n$.
\ENDFOR

\IF{$\sum_{n \in \textbf{N}}\sum_{m \in \textbf{M}_n}{\beta}_m^np_n > p^{\mbox{max}}$}
\STATE Construct matrix \textbf{EEM}.
\REPEAT
\STATE Select an un-marked tuple $(a, n)$ from matrix \textbf{EEM}, which
has the highest non-zero value.
\IF{$a \in [1, \lfloor{p^{\mbox{max}}/{\Delta}}\rfloor]$}
\STATE Assign $a\Delta$ level of power to subchannel $n$.
\ELSE
\STATE Assign $p^{\mbox{max}}$ level of power to subchannel $n$.
\ENDIF
\STATE Mark tuple $(a,n)$ in matrix \textbf{EEM}.
\UNTIL{$\sum_{n \in \textbf{N}}\sum_{m \in {\textbf{M}}_n}{\beta}_m^np_n  = p^{\mbox{max}}$}
\ENDIF
\end{algorithmic}
\end{algorithm}

\noindent
\textbf{Computational complexity of \textit{Algorithm
\ref{alg:sopt-pwr-alloc}}:} Ideally, the GP-based single condensation method is supposed to adopt a conventional optimization solver, such as the interior-point method to solve the transformed convex problem in each iteration. Since the number of iterations is constant for the single condensation method to solve the problem in (\ref{eq:sopt-GP1})-(\ref{eq:sopt-GP2}), its computational complexity is equivalent to the complexity of the adopted convex optimization solver. In~\cite{YNesterov1994}, it is shown that the interior-point method can obtain the $\epsilon$-optimal solution in polynomial time, and hence we can say that the complexity of the single condensation method to solve the problem in (\ref{eq:sopt-GP1})-(\ref{eq:sopt-GP2}) has a polynomial time complexity. Moreover, although the off-the-shelf GP solvers do not solve a GP problem in a native manner, since the number of variable(s) of the problem in (\ref{eq:sopt-GP1})-(\ref{eq:sopt-GP2}) is one, the computational complexity of the resultant solution is constant and quick unlike that in Section~\ref{sssec:GP-joint}. The steps in between $1$ and  $3$, the problem in~(\ref{eq:sopt-GP1})-(\ref{eq:sopt-GP2}) needs to be solved $N$ times. If the resultant total power does not satisfy the power constraint, we need to construct matrix \textbf{EEM}. Except the $1$st row, for each tuple of matrix
\textbf{EEM}, the problem in (\ref{eq:sopt-GP1})-(\ref{eq:sopt-GP2}) needs to be solved twice while replacing $p^{\mbox{max}}$ by different values. Therefore, the process of constructing matrix \textbf{EEM} takes a polynomial time multiplied with $(\lfloor{p^{\mbox{max}}/{\Delta}}\rfloor+1) \times N$. Since each tuple is marked once it is selected for the assignment of power, the steps in between $6$ and $14$ run at most $(\lfloor{p^{\mbox{max}}/{\Delta}}\rfloor+1) \times N$ times. As a result, the complexity of constructing matrix \textbf{EEM} is dominated by the rest of the other steps in \textit{Algorithm
\ref{alg:sopt-pwr-alloc}}.

%After obtaining the subchannel-user mapping information $\textbf{M}_n, n \in \textbf{N}$ using \textit{Algorithm~\ref{alg:sc-usr-map}}, since the number of degrees of freedom of the problem in~(\ref{eq:sopt-indiv}) becomes quite less, the computational complexity of this algorithm is much lower than the approach in Section \ref{sssec:GP-joint} for the given number of subchannels and users.

%   Since the problem in~(\ref{eq:sopt-indiv}) is convex, the solution of this problem has polynomial time complexity. 

 \section{Performance Evaluation}
 \label{sec:eval}

In this section, via simulation, we evaluate the performance of our proposed energy-efficient downlink resource allocation schemes. Followed by the methodology, we exhibit the detailed outcome of the simulation in order to verify the effectiveness of the proposed schemes.

 \subsection{Simulation Setup}
 
The cellular network, that we consider, is isolated from the neighboring networks. It has a circular-like shape and suitable for an office environment. We place the BS at the center of the network, and the users are uniformly spread surrounding the BS within $500$ m distance. Typically, the minimum relative distance between two emploees in an office environment could be $40$ m, and hence we set the minimum distance between two users as $40$ m. Although we set a minimum relative distance between two users, the performance of each NOMA user is independent of this distance~\cite{MBasit2018}. We also set the minimum distance from each user to the BS as $50$ m. As mentioned previously, continuous time is divided into frames. Each time frame is equivalent to $1$ s. During each time frame, the designated spectrum is equally subdivided among $20$ subchannels, and these subchannels are available to be allocated among $M$ users in the system. Each subchannel is assumed to have $200$ KHz bandwidth. According to~\cite{XQiu1999}, the theoretical limit of the channel capacity is given by $\frac{-1.5}{\mbox{ln}(5P_b)}$, where $P_b$ denotes the Bit Error Rate (BER). The BER for each subchannel is configured as $10^{-6}$. We consider that the shadow and Rayleigh are two main fading components of a wireless channel between the BS and a user in the system. The shadow fading follows the log-normal distribution with variance $3.76$.

In order to calculate the quantity of the shadow fading over a subchannel, we assume the reference distance as $1$ km and the SNR for this reference distance is $28$ dB. The Rayleigh fading for all users over all subchannels follows the Rayleigh distribution with $0$ mean and $4.3$ variance. Using all these parameters, the gain of each subchannel for a user towards the BS is computed following ($22$) in~\cite{RRuby2015}. We employ the SIC technique in~\cite{SVanka2012, NIMiridakis2013} for a user in order to decode the SC-coded signal of each subchannel transmitted by the BS.

In addition to implementing our proposed resource allocation schemes, we have implemented the scheme in~\cite{FFang2016}. In the following, this scheme is referred as ``Scheme in~\cite{FFang2016}''. In order to show that the many-to-many matching model applied in the subchannel-user mapping algorithm outperforms the one-to-many matching model, we have one version (i.e., Proposed Scheme-1) which uses \textit{Algorithm~\ref{alg:sc-usr-map}} to solve the subchannel-user mapping problem and the DC programming-based approach for the final power allocation across the subchannels and users. Note that in this version, for step $14$ of \textit{Algorithm~\ref{alg:sc-usr-map}}, ($18$)~\cite{FFang2016} is used. On the other hand, in order to show the superiority of our fine-grained power allocation scheme via the GP technique over the DC programming-based power allocation approach, we have another version (i.e., Proposed Scheme-$2$). This version adopts the subchannel-user mapping algorithm in~\cite{FFang2016} for the first step, and adopts the scheme in Section~\ref{sssec:GP-joint} in order to have power allocation across each allocated subchannel-user slot. We have three more versions of our proposed schemes, namely Scheme-$3$, Scheme-$4$ and Scheme-$5$, in order to exhibit a tradeoff between optimality and computational complexity. Both Scheme-$3$ and Scheme-$4$ adopt the GP technique in Section~\ref{sssec:GP-joint} for the final power allocation. However, Scheme-$3$ uses ($18$)~\cite{FFang2016} for step $14$ in \textit{Algorithm~\ref{alg:sc-usr-map}}, and Scheme-$4$ uses the GP technique in Section~\ref{sssec:GP-indiv}. Consequently, step $14$ in \textit{Algorithm~\ref{alg:sc-usr-map}} happens in constant time for Scheme-$3$ version. On the other hand, this operation of Scheme-$4$ has polynomial time complexity w.r.t. $\textbf{M}_n$. Since the maximum value of $\textbf{M}_n$ is $K$ (e.g., $4$ in our simulation), this complexity of Scheme-$4$ can be considered negligible, but provides better solution compared to Scheme-$3$. Moreover, our Scheme-$5$ version uses the technique in Section~\ref{sssec:GP-indiv} for step $14$ of the subchannel-user mapping algorithm, and uses \textit{Algorithm 2} for the final fine-grained power allocation across the subchannels and users. The complexity of the suboptimal power allocation algorithm (i.e., \textit{Algorithm 2}) is elaborately explained in Section~\ref{sssec:sopt-joint}. Apparently, both \textit{Algorithm 2} and the fine-grained power allocation scheme in Section~\ref{sssec:GP-joint} have polynomial time complexity. However, since the complexity of the GP technique is computationally intensive with the large number of optimization variables (e.g., $\sum_{n \in \textbf{N}}|\textbf{M}_n|$), the proposed Scheme-$3$ or Scheme-$4$ may run very slowly with the growing number of users and subchannels in the system. On the other hand, although step $2$ of \textit{Algorithm 2} is associated with solving the problem in~(\ref{eq:sopt-GP1})-(\ref{eq:sopt-GP2}) via the GP technique, the corresponding problem has only one optimization variable which results in constant time complexity for this step. Other steps of \textit{Algorithm 2} have constant time complexity, and hence the complexity of Scheme-$5$ can be considered negligible compared to Scheme-$3$ or Scheme-$4$. However, because of the intense search of the fine-grained power allocation scheme in Section~\ref{sssec:GP-joint}, Scheme-$3$ and Scheme-$4$ (especially Scheme-$4$) achieve much better performance compared to Scheme-$5$, which has been verified in the subsequent discussions. Finally, in order to demonstrate the global optimal performance, we apply the brute-force search on both the subchannel-user mapping and power allocation problems for a system with $10$ and $20$ users. In the following subsection, for each data point, we conduct the simulation over $10000$ times.

 \subsection{Simulation Results}
 
% \begin{figure*}%
%\centering
%\begin{subfigure}{0.7\columnwidth}
%\includegraphics[width=\columnwidth]{figures/syseff_usrs.eps}%
%\caption{With the increasing number of users.}%
%\label{fig:syseff-usrs}%
%\end{subfigure}%
%\begin{subfigure}{0.7\columnwidth}
%\includegraphics[width=\columnwidth]{figures/syseff_pwr.eps}%
%\caption{With the increasing total power at the BS.}%
%\label{fig:syseff-pwr}%
%\end{subfigure}%
%\begin{subfigure}{0.7\columnwidth}
%\includegraphics[width=\columnwidth]{figures/syseff_pc.eps}%
%\caption{With the increasing circuit power consumption.}%
%\label{fig:syseff-pc}%
%\end{subfigure}%
%\caption{Comparison of overall energy efficiency among different schemes.}
%\label{figabc}
%\end{figure*}

\begin{figure}
  \begin{center}
    \includegraphics[width=0.5\columnwidth]{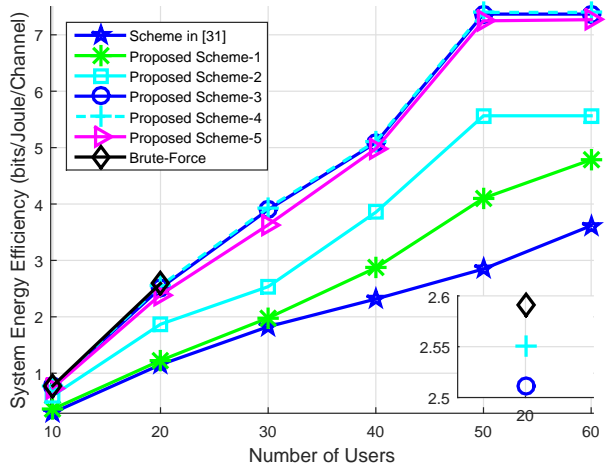}
    \caption{Comparison of overall energy efficiency with the increasing number of users.}
    \label{fig:syseff-usrs}
  \end{center}
\end{figure}

%   As we argued previously that many-to-many matching model is able to exploit the multi-user-channel diversity of wireless systems well compared to the one-to-many matching one

%  This is because the more the diversity of multi-user-channel is exploited, the larger the sum-rate is, given some constant power level at the system.

\begin{figure}[h!]%
    \centering
    \subfloat[Comparison of allocated users to each individual subchannel.\label{fig:usrs-vs-sc}]{{\includegraphics[scale=0.6]{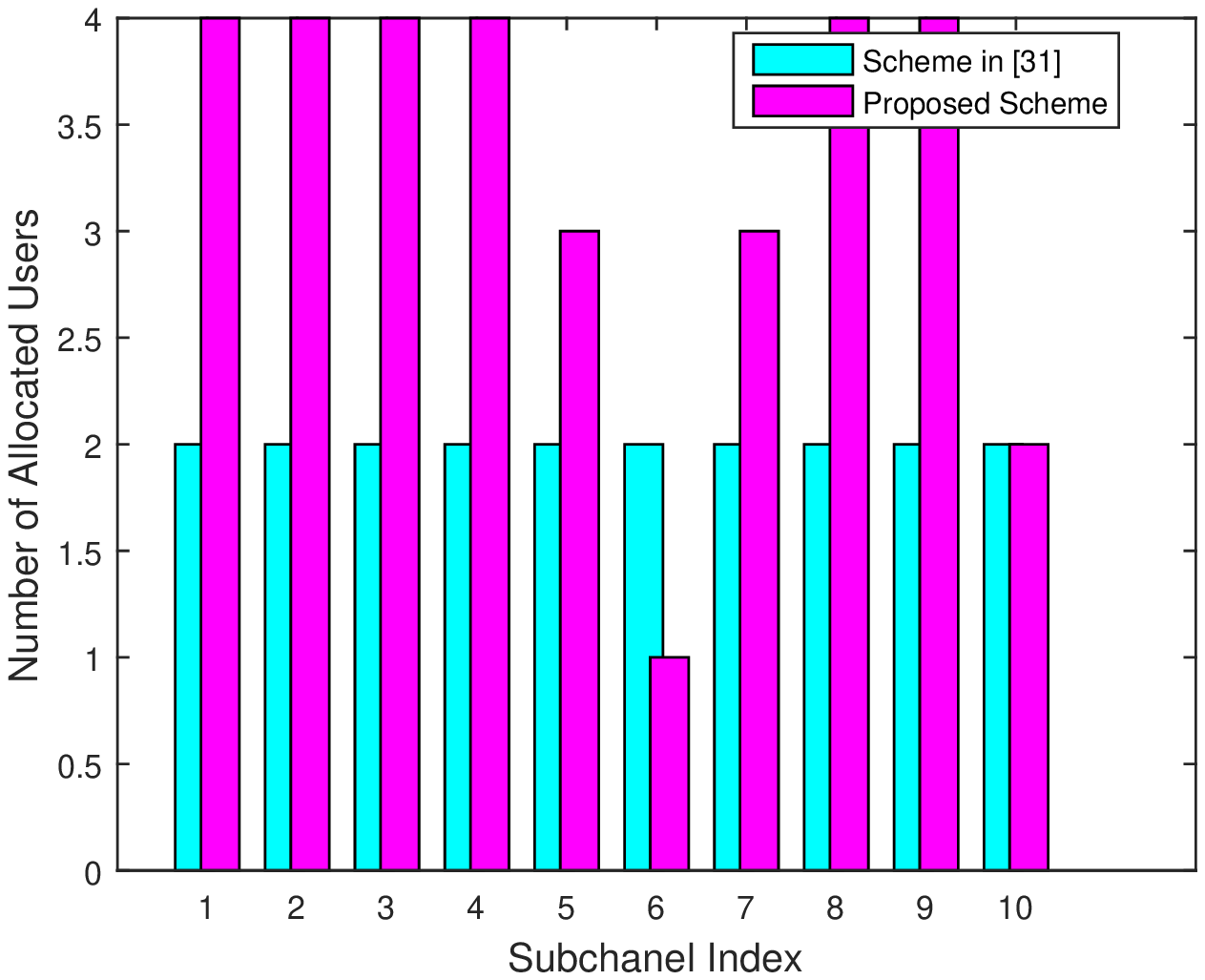} }}%
   ~
    \subfloat[Comparison of allocated subchannels to each individual user.\label{fig:scs-vs-usr}]{{\includegraphics[scale=0.6]{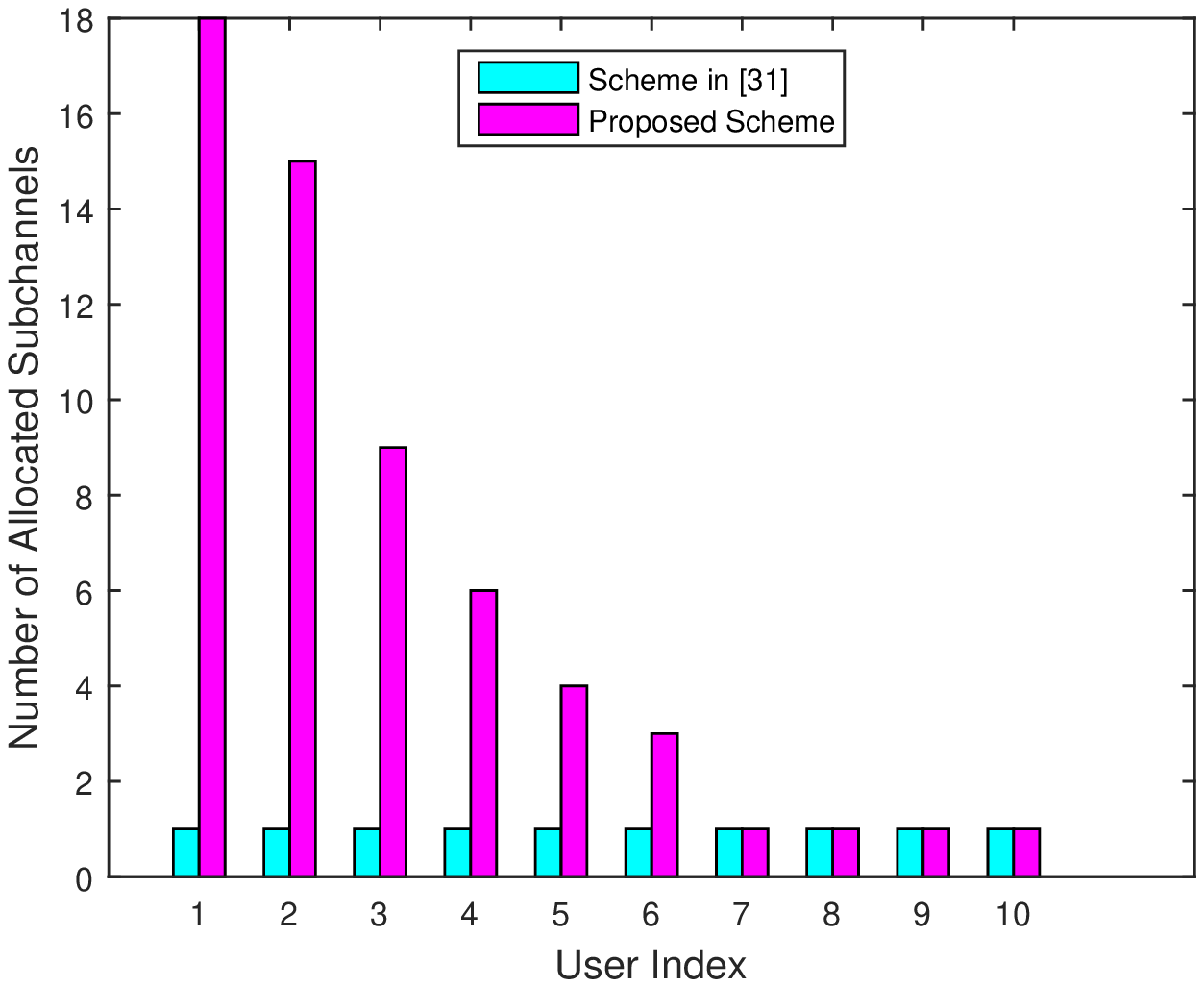} }}%
    \caption{Detailed subchannel-user assignment study when $M=40$.}%
    %\label{fig:convg}%
\end{figure}

For $K = 4$, $p^{\mbox{max}} = 23~\mbox{dBW}$ and $p_c = 1.75~\mbox{dBW}$, Fig.~\ref{fig:syseff-usrs} presents the increasing energy efficiency with the increasing number of users. Given the constant power level at the BS, the more the users, the higher the sum-rate, and hence the larger the efficiency is. This trend is similar for all the schemes no matter it is ours or the scheme in~\cite{FFang2016}. Via the many-to-many matching model, if a user has better gain over many subchannels compared to other users, that user can be assigned to as many subchannels as possible if such assignments increase the system energy efficiency. On the other hand, via the one-to-many matching model, it is possible that a user with worse gain is forced to be assigned to a subchannel, despite such assignment is not necessarily beneficial for the system energy efficiency. This is because each user can be assigned to at most one subchannel via this matching model. As a result, both intuitively and empirically, we see the evidence of enhanced performance for our Scheme-$1$ compared to that in~\cite{FFang2016}, even when we use (18)~\cite{FFang2016} for step $14$ in \textit{Algorithm~\ref{alg:sc-usr-map}} and the DC programming-based approach for the final power allocation. To provide further evidence, for $M = 40$, we plot Fig.~\ref{fig:usrs-vs-sc} and Fig.~\ref{fig:scs-vs-usr}, which are the outcome of the subchannel-user mapping algorithms. In Fig.~\ref{fig:scs-vs-usr}, we compare the number of allocated subchannels to each individual user (the users are sorted in the ascending order of their distance from the BS) between the one-to-many and our model. From our observation, we see that via the many-to-many matching model, a user can be assigned to multiple subchannels, and this number depends on the subchannels over which that particular user can achieve enhanced energy efficiency compared to other users. Our channel model is such that the quantity of the shadow fading is dominated by that of the Rayleigh fading. Therefore, a user that is closer to the BS is likely to have better gain over more subchannels compared to a farther user. Therefore, in Fig.~\ref{fig:scs-vs-usr}, we see that a user that is the closest to the BS is assigned to the largest number of subchannels. On the other hand, based on the dynamics of users' gain, in order to enhance the energy efficiency further, some subchannels accommodate more users compared to other subchannels.

%if a subchannel has better gain for more users compared to other subchannels, it is likely that this subchannel is assigned with more users in order to enhance the overall energy efficiency of the system.  

%    this user can  
%of users each subchannel can afford (i.e.,  a subchannel with better gain can accommodate more users compared to others). Moreover, it is likely that the better the channel condition a user has, can obtain more subchannels.} 

\begin{comment}

\begin{figure}
  \begin{center}
    \includegraphics[width=0.5\columnwidth]{usrs_vs_sc.eps}
    \caption{Comparison of allocated users to each individual subchannel when $M=40$.}
    \label{fig:usrs-vs-sc}
  \end{center}
\end{figure}

\begin{figure}
  \begin{center}
    \includegraphics[width=0.5\columnwidth]{scs_vs_usr.eps}
    \caption{Comparison of allocated subchannels to each individual user when $M=40$.}
    \label{fig:scs-vs-usr}
  \end{center}
\end{figure}
\end{comment}

In Fig.~\ref{fig:usrs-vs-sc}, we compare the number of assigned users to each individual subchannel between two subchanel-user mapping algorithms. In this figure, although $K=4$, not necessarily all subchannels have $4$ allocated users. In general, it is seen that if the difference of gain between any two assigned users over any subchannel is larger compared to other subchannels, that subchannel achieves better sum-rate as well as better energy efficiency. The numerical degree of users' gain over a subchannel also plays a crucial role in the decision whether additional user will be allocated to that subchannel or not. Because of these user dynamics, some subchannels cannot be assigned to many users as the allocation of more and more users may decrease the sum-rate as well as its energy efficiency, which is proved in \textit{Proposition 1}. Due to the structure of the formulation, unlike the DC programming-based approach, the GP technique can provide fine-grained energy-efficient power allocation across all subchannel-user tuples. Hence, the energy efficiency is much better for this case (i.e., Scheme-$3$) even if we use (18)~\cite{FFang2016} for step $14$ of \textit{Algorithm~\ref{alg:sc-usr-map}}. If we use the GP technique instead of (18)~\cite{FFang2016} for step $14$ in \textit{Algorithm~\ref{alg:sc-usr-map}} (i.e., Scheme-$4$), we obtain better organization in the subchannel-user map due to the better power allocation among the users of each subchannel. Therefore, this scheme has the best performance compared to all the others. The elegance of the fine-grained GP technique is further evident from the enhanced performance of Scheme-$2$ over the scheme in \cite{FFang2016}. In Scheme-$2$, we take the same subchannel-user mapping algorithm (via the one-to-many matching model) as that in \cite{FFang2016}. On the other hand, we argued previously that the fine-grained power allocation using the GP technique is computationally quite intensive. Consequently, we developed \textit{Algorithm~\ref{alg:sopt-pwr-alloc}} in order to have fine-grained energy-efficient power allocation across all the allocated subchannel-user tuples in a low complexity manner. Because of adopting the insights of the optimal solution, the resource allocation scheme via this algorithm (i.e., Scheme-$5$) has very close performance to that of the Scheme-$3$ and Scheme-$4$ versions, and obviously outperforms the scheme with the DC programming-based approach.

% With our subchannel-user mapping algorithm, even if we use DC programming for the final power allocation, we see the enhanced performance of our scheme (i.e., Scheme-$1$) compared to that in~\cite{FFang2016}

\begin{figure}
  \begin{center}
    \includegraphics[width=0.5\columnwidth]{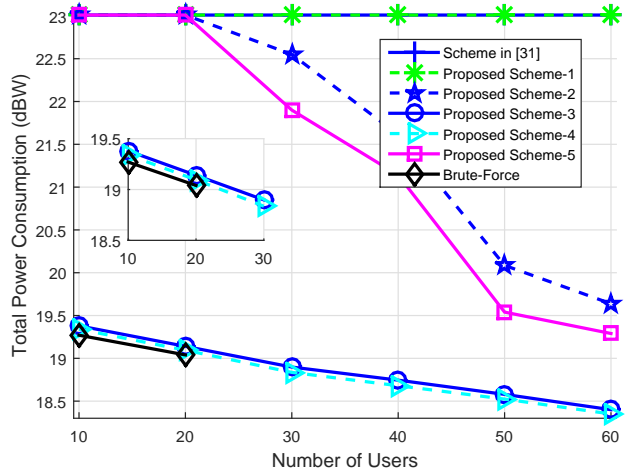}
    \caption{Comparison of total power consumption with the increasing number of users.}
    \label{fig:syspwr-usrs}
  \end{center}
\end{figure}

Given $p^{\mbox{max}} = 23~\mbox{dBW}$ and $p_c = 1.75~\mbox{dBW}$, Fig.~\ref{fig:syspwr-usrs} presents total power consumption with the increasing number of users. This is natural that the more the users in the system, the diversity of the users that spread among all the allocated subchannels (i.e., multi-user diversity) increases. Therefore, with the increasing number of users, it is more likely that each subchannel is assigned to at least one user with good channel. On the other hand, from our observation, it is seen that if a subchannel has at least one user with good channel, it requires less power for that subchannel to reach the optimal energy-efficient state. Consequently, a lower power level should be required to reach the optimal energy-efficient state for a system with more users compared to that with less users. This observation and insights hold for both the GP technique and \textit{Algorithm~\ref{alg:sopt-pwr-alloc}}. However, since the mechanism of \textit{Algorithm~\ref{alg:sopt-pwr-alloc}} is somewhat suboptimal (although developed based on the insights of the optimal solution) compared to that in Section~\ref{sssec:GP-joint}, total consumed power in this case is slightly larger compared to the other one. The findings of this figure are quite interesting in a sense that the schemes with the fine-grained power allocation via the GP technique and \textit{Algorithm~\ref{alg:sopt-pwr-alloc}} use much less power compared to that with the DC programming-based approach. If we compare Fig.~\ref{fig:syseff-usrs} with this figure, it becomes even more interesting as our schemes achieve much better energy efficiency using much less power compared to that with the DC programming-based approach. On the other hand, the schemes with the DC programming-based approach use full power of the BS, but incur much less energy efficiency. From the perspective of green communications, the results presented in this figure verify our original motivation towards pursuing this work.

% Nonetheless, Shannon’s information capacity theorem illustrates that the  Note that under the consideration of fixed circuit power, there always exists an optimal point in the energy efficiency versus spectrum efficiency curve.

\begin{figure}
  \begin{center}
    \includegraphics[width=0.5\columnwidth]{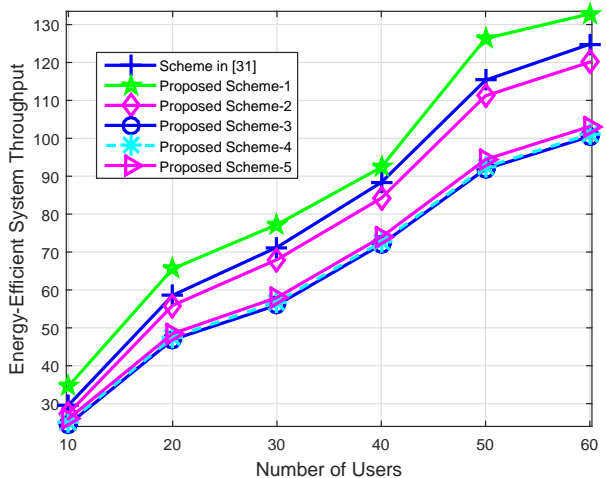}
    \caption{Comparison of energy-efficient total throughput with the increasing number of users.}
    \label{fig:systhr-usrs}
  \end{center}
\end{figure}

Similar to our work, the conventional definition of energy efficiency is achievable blocks of bits from a channel under the usage of unit power level. The Shannon's information capacity theorem has already established that two objectives, i.e., minimizing the consumed energy and maximizing the spectral efficiency are not achievable simultaneously at the same operating point. Consequently, under the consideration of fixed circuit power, there always exist two separate optimal points in the energy efficiency versus spectrum efficiency curve.  In Fig.~\ref{fig:systhr-usrs}, we compare the energy-efficient total throughput acheievd by all aforementioned schemes. The more the users, the better the utilization of limited resources because of the enhanced multi-user diversity. Hence, we see the increasing trend in the energy-efficient total throughput, achieved by all the schemes, with the increasing number of users. However, due to the fact in the Shannon's information capacity theorem, since the scheme in~\cite{FFang2016} does not achieve the optimal energy-efficient state, it is possible that the total energy-efficient throughput achieved by this scheme is larger than our schemes. This is what observed in this figure.

\begin{figure}[h!]%
    \centering
    \subfloat[Overall energy efficiency.\label{fig:syseff-pwr}]{{\includegraphics[scale=0.6]{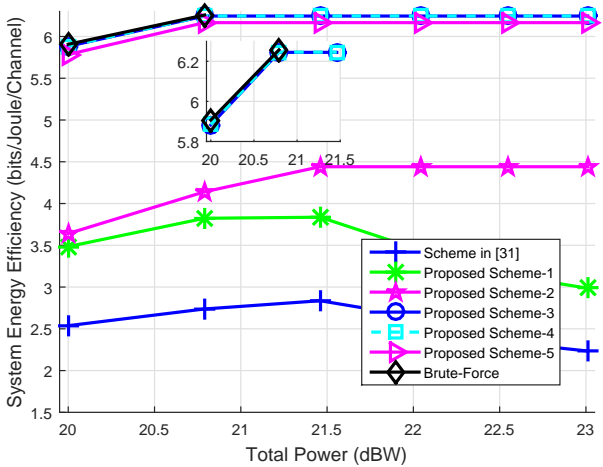} }}%
   ~
    \subfloat[Total power consumption.\label{fig:syspwr-pwr}]{{\includegraphics[scale=0.6]{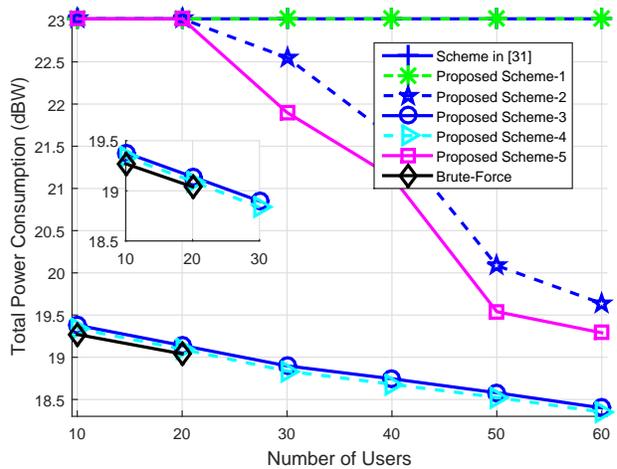} }}%
    \caption{Comparison of overall energy efficiency and total power consumption with the increasing power at the BS.}%
    %\label{fig:convg}%
\end{figure}

In Fig.~\ref{fig:syseff-pwr}, given $M = 40$ and $p_c = 1.75~\mbox{dBW}$, we show the energy efficiency with the increasing power at the BS. As we saw in the previous results that the scheme in~\cite{FFang2016} fails to exploit the multi-user diversity of wireless systems as well as uses the DC programming-based approach to obtain the coarse-grained power allocation, this scheme has the lowest energy efficiency no matter the power constraint of the BS is. For this case, we see that with the increasing power level, the trend of energy efficiency is decreasing. This is because there is a tradeoff between the transmission capacity and the energy-efficient power consumption. Whereas, for our case, since the GP technique provides a fine-grained elegant power allocation, the overall energy efficiency is much better compared to the benchmark scheme. Moreover, since the GP technique provides the unique solution while consuming much less power, no matter we increase the power level of the BS, the energy efficiency remains same at the unique point. Similar trend is observed in the case of our suboptimal \textit{Algorithm~\ref{alg:sopt-pwr-alloc}} although the energy efficiency achieved by this scheme is slightly lower compared to the fine-grained GP-based power allocation scheme. In order to show the power consumption for this case, we plot Fig.~\ref{fig:syspwr-pwr}. Since the optimal energy-efficient state of the system is unique and our proposed GP technique is able to search this state to some extent, we see the constant level of used power no matter how much power the BS has. In the similar manner, given the subchannel-user mapping matrix, via our suboptimal \textit{Algorithm~\ref{alg:sopt-pwr-alloc}}, we obtain an elegant energy-efficient point for each individual subchannel. Consequently, although this suboptimal scheme does not achieve as good solution as by that in Section~\ref{sssec:GP-joint}, the suboptimal unique energy efficiency is still achieved at the unique power level. Whereas, the schemes, which use the DC programming-based approach, consume full power of the BS. Therefore, in this case, we see the increasing power consumption with the increasing total power at the BS.

%\begin{figure}
%  \begin{center}
%    \includegraphics[width=0.8\columnwidth]{figures/syseff_pc.eps}
%    \caption{Comparison of overall energy efficiency with the increasing level of circuit power.}
%    \label{fig:syseff-pc}
%  \end{center}
%\end{figure}

\begin{comment}
\begin{figure}
  \begin{center}
    \includegraphics[width=0.5\columnwidth]{figures/syseff_pc.eps}
    \caption{Comparison of overall energy efficiency with the increasing $p_c$.}
    \label{fig:syseff-pc}
  \end{center}
\end{figure}

\begin{figure}
  \begin{center}
    \includegraphics[width=0.5\columnwidth]{figures/syspwr_pc.eps}
    \caption{Comparison of total power consumption with the increasing $p_c$.}
    \label{fig:syspwr-pc}
  \end{center}
\end{figure}
\end{comment}

\begin{figure}[h!]%
    \centering
    \subfloat[Overall energy efficiency.\label{fig:syseff-pc}]{{\includegraphics[scale=0.6]{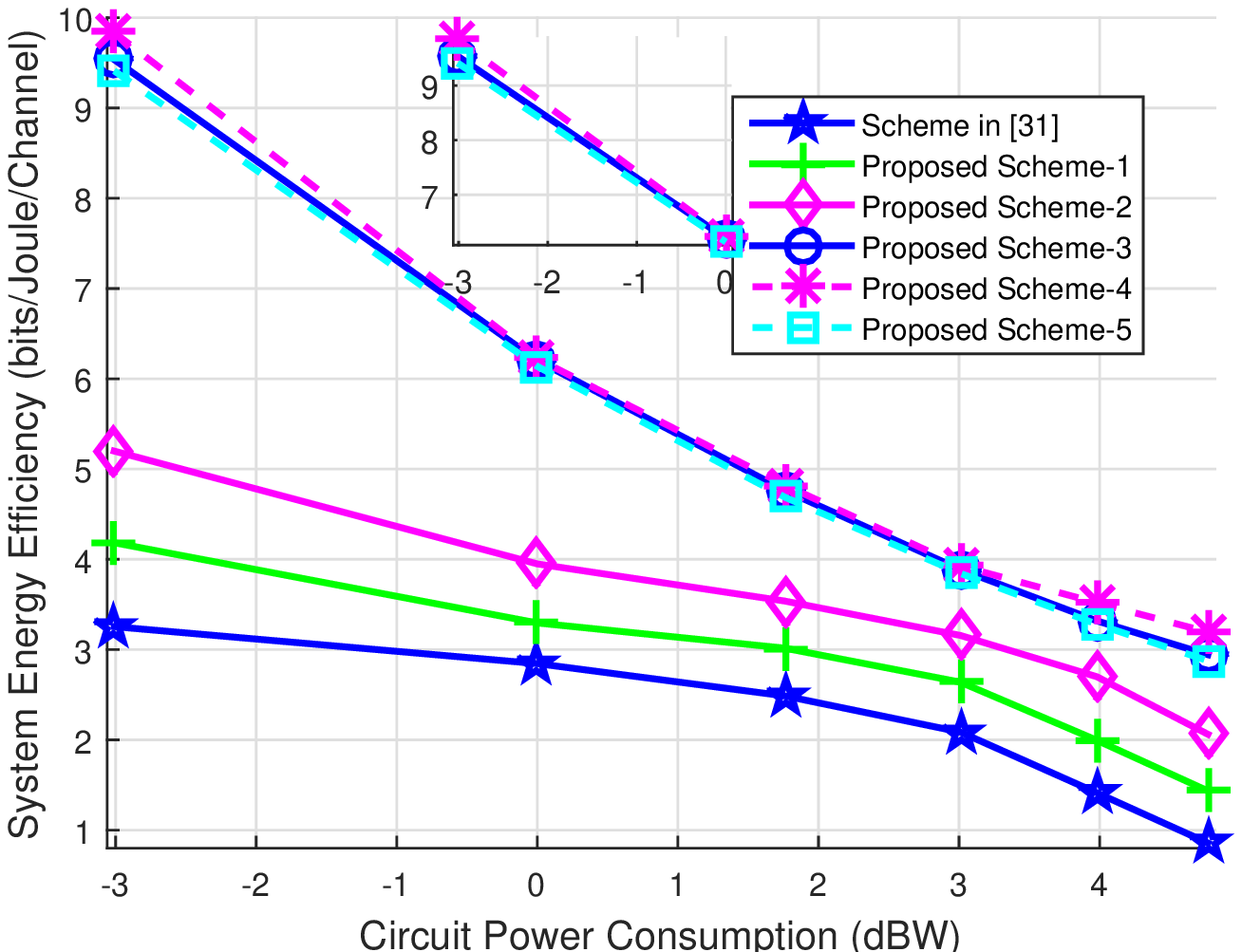} }}%
   ~
    \subfloat[Total power consumption.\label{fig:syspwr-pc}]{{\includegraphics[scale=0.6]{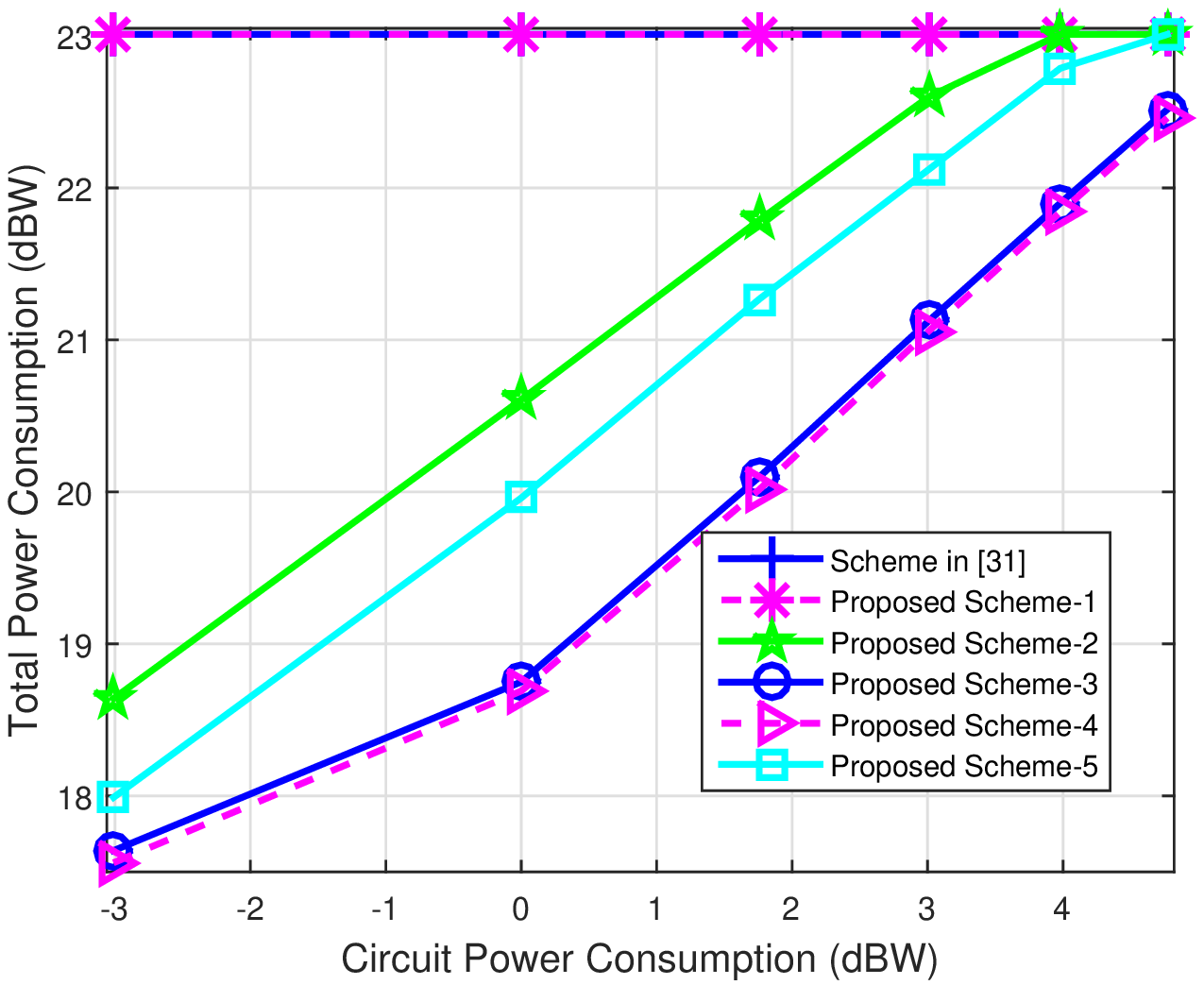} }}%
    \caption{Comparison of overall energy efficiency and total power consumption with the increasing $p_c$.}%
    %\label{fig:convg}%
\end{figure}

In Fig.~\ref{fig:syseff-pc}, given $p^{\mbox{max}} = 23~\mbox{dBW}$ and $M = 40$, we show the decreasing overall energy efficiency with the increasing circuit power consumption. This is a natural trend achieved by all the schemes. Since circuit power is not used in enhancing system throughput but is added as consumed power, the overall energy efficiency is decreasing with the increasing circuit power. However, our subchannel-user mapping scheme can better exploit the multi-user diversity of wireless systems and we use the GP technique for the final fine-grained energy-efficient power allocation, our schemes, even via our suboptimal \textit{Algorithm~\ref{alg:sopt-pwr-alloc}}, always outperform the scheme in~\cite{FFang2016}. In order to show the total power consumption in this case, we plot Fig.~\ref{fig:syspwr-pc} with the increasing $p_c$. Increasing $p_c$ means, a subchannel requires higher power to reach its energy-efficient state. Consequently, the optimal energy efficiency is achieved at larger power level with the increasing $p_c$. For the similar reason, via our suboptimal \textit{Algorithm~\ref{alg:sopt-pwr-alloc}}, the same trend is observed with the increasing $p_c$. On the other hand, since the DC programming-based approach uses full power of the BS no matter the value of $p_c$ is, the total power consumption is constant (i.e., $p^{\mbox{max}}$) via the scheme in~\cite{FFang2016} and our Scheme-$1$ version.    

\section{Conclusion}
\label{sec:concl}

In this paper, we proposed an energy-efficient downlink subchannel and power allocation scheme for NOMA systems with enhanced performance compared to the most relevant existing work in the literature. Due to the discrete nature of subchannel assignment and the characteristics of the NOMA technique, this is an MINLP problem. Therefore, similar to an existing work, we solved the problem via decomposing it into a subchannel allocation subproblem followed by a power loading subproblem. However, unlike the existing work, via a many-to-many matching model, we better exploited the multi-user diversity of wireless systems in the solution of the subchannel-user mapping subproblem. In the second step, unlike the DC programming-based approach, via the GP technique, we were able to allocate the power level across the allocated subchannel-user slots in a fine-grained manner such that better energy efficiency is achieved compared to the benchmark scheme. Since the fine-grained power allocation via the GP technique is computationally intensive using the off-the-shelf GP solvers, we also proposed a suboptimal fine-grained power allocation algorithm with much lower computational complexity. Under various realistic scenarios, extensive simulation had been conducted to verify that our scheme (even via our computationally-efficient suboptimal power allocation algorithm) can outperform the existing scheme while consuming much less power in the system.

Besides achieving an elegant energy-efficient state via the better resource allocation schemes, the implication of this work is extended to a certain extent. Via our schemes (even the suboptimal one), since better energy efficiency is achieved at a lower power level, the interference effect to the neighboring networks is expected to be mitigated. At the same time, unused power in the system can be used for other purposes.

\begin{appendices}

\section{Proof of Proposition 1}

\begin{figure}
  \begin{center}
    \includegraphics[width=0.5\columnwidth]{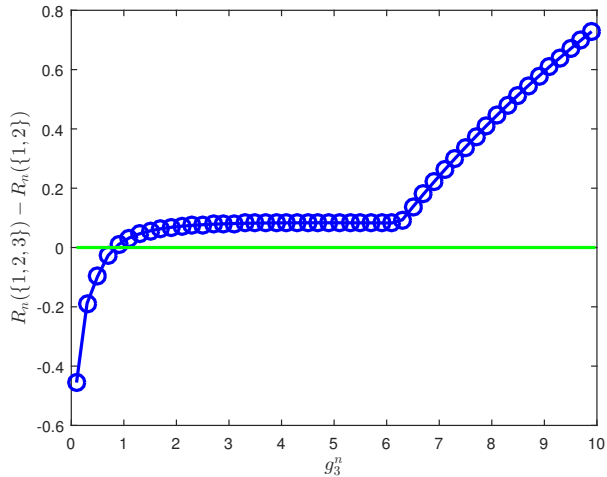}
    \caption{A sample example that $[R_n(\{1,2,3\})-R_n(\{1,2\})]$ can be both positive and negative.}
    \label{fig:prop1-proof}
  \end{center}
\end{figure}

Consider a subchannel $n$, and it has two allocated users, indexed by $1$ and $2$. Moreover, we set $K = 4$, and there is one assumption, i.e., $R_n(\{1, 2\}) > R_n(\{1\})$. It implies that the sum-rate of user $1$ and user $2$ is larger than that of only user $1$. Now, user $3$ has come to be assigned with subchannel $n$. There are two possible conditions for this assignment, which are $R_n(\{1,2,3\}) > R_n(\{1,2\})$ and $R_n(\{1,2,3\}) < R_n(\{1,2\})$. According to step $15$ of \textit{Algorithm~\ref{alg:sc-usr-map}}, we only consider this user to construct an addition strategy if and only if the former case is true. In practice, both the conditions for any subchannel $n$ can happen. This statement can be proved from the result of Fig.~\ref{fig:prop1-proof}. In this figure, we plot $R_n(\{1,2,3\}) - R_n(\{1,2\})$ w.r.t. $g_3^n$. It is obvious that $R_n(\{1,2,3\}) - R_n(\{1,2\})$ can be both positive and negative. The energy efficiency of subchannel $n$ is enhanced if and only if the aforementioned value is positive, and user $3$ is not added to subchannel $n$ for the negative case. This completes the proof. Note that $g_1^n = 0.4141,~g_2^n = 6.2512$ and $p_n = 50$ in Fig.~\ref{fig:prop1-proof}. 

\section{Proof of Proposition 2}

\begin{figure}
  \begin{center}
    \includegraphics[width=0.5\columnwidth]{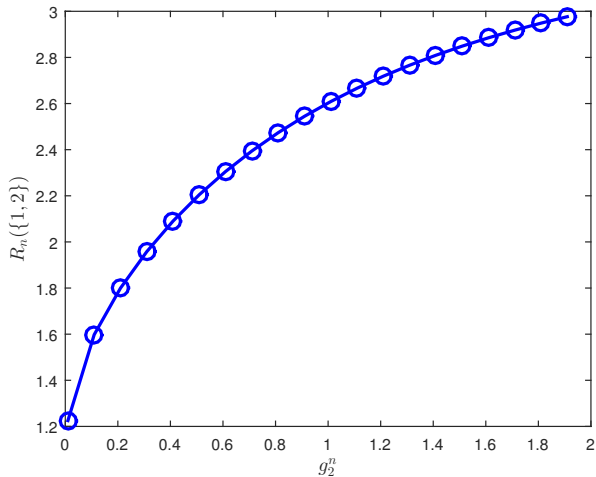}
    \caption{A sample example that $R_n(\{1,2\})$ is increasing w.r.t. $g_2^n$.}
    \label{fig:prop2-proof}
  \end{center}
\end{figure}

Without loss of generality, let us assume $K=2$. Consider that subchannel $n$ has already $2$ users, indexed by $1$ and $2$, and $g_1^n > g_2^n$ holds. At this point, user $3$ has come to be assigned with subchannel $n$ with $g_3^n > g_2^n~\mbox{and}~g_1^n > g_3^n$. Since the nature of function $R_n(\{1,2\})$ is increasing w.r.t. $g_2^n$ (as shown in Fig.~\ref{fig:prop2-proof}), we can conclude $R_n(\{1,3\}) > R_n(\{1,2\})$. Moreover, using (18)~\cite{FFang2016}
and then simplifying, $R_n(\{1,3\})$ and $R_n(\{3,2\})$ are given by

\begin{eqnarray}
\label{prop2:f}  & R_n(\{1,3\}) = \\
& \nonumber \mbox{log}_2\left((1+\frac{p_n(g_1^n)^{1+\alpha}}{(g_1^n)^{\alpha}+(g_3^n)^{\alpha}})(1
+ \frac{p_n(g_1^n)^{1+\alpha}}{(g_1^n)^{\alpha}+(g_3^n)^{\alpha}+p_ng_3^n(g_1^n)^{\alpha}})\right),
\\
\label{prop2:l} & R_n(\{3,2\}) = \\
& \nonumber \mbox{log}_2\left((1+\frac{p_n(g_3^n)^{1+\alpha}}{(g_2^n)^{\alpha}+(g_3^n)^{\alpha}})(1
+ \frac{p_n(g_2^n)^{1+\alpha}}{(g_2^n)^{\alpha}+(g_3^n)^{\alpha}+p_ng_2^n(g_3^n)^{\alpha}})\right),~\mbox{respectively}.
\end{eqnarray}

Since $g_1^n > g_3^n > g_2^n$, from (\ref{prop2:f}) and
(\ref{prop2:l}), for any $\alpha < 0$, it is straightforward to say $R_n(\{1,3\}) > R_n(\{3,2\})$. Now, if another user $4$ comes, two possible cases are possible. The first case is, user $4$ will replace further user $3$ if both $g_4^n > g_3^n~\mbox{and}~g_1^n > g_4^n$ hold. In order to prove this proposition, we need to show that rejected user $2$ cannot match with subchannel $n$ further as the assignment of user $2$ to subchannel $n$ cannot enhance the energy efficiency any more. For this case, since $g_3^n > g_2^n$, via substituting user $3$ by user $4$ in (\ref{prop2:f}) and (\ref{prop2:l}), $R_n(\{1,4\}) > R_n(\{2,4\})$ can be
proved. For the sake of soundness, the second possible case is, user $4$ approaches subchannel $n$ with $g_4^n < g_3^n$. In this case, user $3$ is not substituted by user $4$ as $R_n(\{1,4\})$ is an increasing function of $g_4^n$ according to Fig.~\ref{fig:prop2-proof}. Now, our objective is to prove $R_n(\{1,3\}) > R_n(\{3,4\}) > R_n(\{2,4\})$. While substituting user $2$ by user $4$ in (\ref{prop2:f}) and (\ref{prop2:l}),
$R_n(\{1,3\}) > R_n(\{3,4\})$ is straightforward as $g_1^n > g_3^n > g_4^n$ is true.  In the similar manner,  $R_n(\{3,4\}) > R_n(\{2,4\})$ can be proved as well due to the $g_3^n > g_4^n$ and $g_3^n > g_2^n$ relations. Thus, the proof of this proposition is completed.

\section{Proof of Theorem 1}

Let us prove this theorem by contradiction. Consider a matching relation $\mu$ and a user-subchannel pair $(m, n)$, which satisfy $m \not\in {\mu}(n)$ and $n \not\in {\mu}(m)$. Although this pair is not matched, both user $m$ and subchannel $n$ prefer each other over the remaining other subchannels and users, respectively. Now, according to the steps of \textit{Algorithm \ref{alg:sc-usr-map}}, two possible cases can happen. The first case is, $|\bm{\Omega}_m| < K$, and user $m$ proposes subchannel $n$ to be matched with (step $5$ in \textit{Algorithm \ref{alg:sc-usr-map}}). Over receiving the proposal, according to \textit{Proposition 1}, if the energy efficiency of subchannel $n$ is enhanced by adding user $m$, subchannel $n$ is paired with user $m$, otherwise not. From the latter case, it can be concluded that although user $m$ prefers subchannel $n$ over the remaining other subchannels, subchannel $n$ does not prefer user $m$ over the remaining other users. The second case is, subchhanel $n$ is already matched with user $m$ through the addition strategy. However, later, in some iteration, user $m'$ proposes subchannel $n$ to be matched with, and subchannel $n$ prefers user $m'$ over user $m$ because of the enhanced energy efficiency. Consequently, user $m$ will be replaced by user $m'$ (step $19$ - step $20$) in ${\mu}(n)$. From this discussion, it can be concluded that if both the user $m$ and subchannel $n$ prefer each other over the remaining other players, there is no way that a matching relation will not be established between them. Consequently, the initial statement of this proof is shown to be false. This concludes the proof of this theorem.

\end{appendices}

\section*{References}

\bibliography{references}

\end{document}